\documentclass[11pt]{article}
\usepackage{amssymb,amsmath,fullpage,graphicx}
\usepackage{enumerate}
\def\x{\mathbf x}

\makeatletter \numberwithin{equation}{section}

\begin{document}
\title{Stability analysis for pitchfork bifurcations of solitary waves in generalized nonlinear Schr\"odinger equations}
\author{Jianke Yang \\
Department of Mathematics and Statistics \\ University of Vermont\\
Burlington, VT 05401, USA}
\date{ }
\maketitle

\begin{abstract}
Linear stability of both sign-definite (positive) and
sign-indefinite solitary waves near pitchfork bifurcations is
analyzed for the generalized nonlinear Schr\"odinger equations with
arbitrary forms of nonlinearity and external potentials in arbitrary
spatial dimensions. Bifurcations of linear-stability eigenvalues
associated with pitchfork bifurcations are analytically calculated.
It is shown that the smooth solution branch switches stability at
the bifurcation point. In addition, the two bifurcated solution
branches and the smooth branch have the opposite (same) stability
when their power slopes have the same (opposite) sign. One unusual
feature on the stability of these pitchfork bifurcations is that the
smooth and bifurcated solution branches can be both stable or both
unstable, which contrasts such bifurcations in finite-dimensional
dynamical systems where the smooth and bifurcated branches generally
have opposite stability. For the special case of positive solitary
waves, stronger and more explicit stability results are also
obtained. It is shown that for positive solitary waves, their linear
stability near a bifurcation point can be read off directly from
their power diagram. Lastly, various numerical examples are
presented, and the numerical results confirm the analytical
predictions both qualitatively and quantitatively.
\end{abstract}



\section{Introduction}
Bifurcation of solitary waves is an important phenomenon in
nonlinear wave equations. One type of bifurcation is the so-called
pitchfork bifurcation, where a smooth branch of solitary waves
exists on both sides of the bifurcation point, but two additional
solution branches bifurcate out to only one side of the bifurcation
point. The most common pitchfork bifurcation of solitary waves is
the symmetry-breaking bifurcation, where solitary waves on the
smooth branch have certain symmetry, but solitary waves on the
bifurcated branches lose that symmetry and become asymmetric. This
symmetry-breaking bifurcation occurs frequently in various nonlinear
wave models originating from diverse physical disciplines (such as
nonlinear optics and Bose-Einstein condensates). For instance, this
bifurcation has been reported in the nonlinear Schr\"odinger (NLS)
equations with external potentials
\cite{Weinstein_2004,Panos_2005_pitchfork,Malomed_2007,Weinstein_2008,Malomed_2008,
Sacchetti_2009,Panos_2009,Kirr_2011,Peli_2011,Akylas_2012,Yang_classification}.
Physically these NLS equations govern nonlinear light propagation in
refractive-index-modulated optical media
\cite{Kivshar_book,Yang_SIAM} and atomic interaction in
Bose-Einstein condensates loaded in magnetic or optical traps (in
the latter community these equations are called the Gross-Pitaevskii
equations \cite{BEC}). This symmetry-breaking bifurcation has also
been reported in the linearly-coupled NLS equations which govern
light transmission in dual-core couplers
\cite{Akhmediev_pitchfork_1993, Malomed_coupler_2011}. Analytical
studies of symmetry-breaking bifurcations have also been made,
mostly for the NLS equations with special types of nonlinearities
and potentials (see
\cite{Weinstein_2004,Weinstein_2008,Sacchetti_2009,Kirr_2011,Peli_2011,Yang_classification}
for instance). In \cite{Weinstein_2004} the authors considered a
one-dimensional NLS equation with focusing cubic nonlinearity and a
Dirac-type symmetric double-well potential, and showed the presence
of symmetry breaking bifurcation as well as the exchange of
dynamical stability from the symmetric branch to the asymmetric
branch at the bifurcation point. In \cite{Weinstein_2008} the
authors considered a class of multi-dimensional NLS equations with
focusing cubic nonlinearity and symmetric potentials, and showed
that symmetry-breaking bifurcation occurs when the power (also
called the squared norm in mathematics and particle numbers in
Bose-Einstein condensation) of the symmetric solitary waves
increases above a certain threshold, provided that the first two
eigenvalues of the linear potential are sufficiently close to each
other (such as in double-well potentials with large separation
between the two wells). In addition, the authors showed that above
this power threshold, the symmetric states become unstable, and a
pair of orbitally stable asymmetric states appear. In
\cite{Sacchetti_2009}, the author considered a class of
multi-dimensional NLS equations with defocusing power nonlinearity
and symmetric double-well potentials in the semiclassical limit, and
showed that symmetry-breaking bifurcations occur for antisymmetric
solitary waves. In \cite{Kirr_2011}, the authors considered a class
of one-dimensional NLS equations with focusing power nonlinearity
and a symmetric potential, and showed that symmetry-breaking
bifurcation occurs for positive symmetric solitary waves if the
potential satisfies certain requirements. In addition, they showed
that the symmetric branch changes stability at the bifurcation
point, and the asymmetric branches can be orbitally stable or
unstable under different conditions. In \cite{Peli_2011}, the
authors considered the same class of equations as in
\cite{Kirr_2011} and obtained normal forms for these
symmetry-breaking bifurcations. In \cite{Yang_classification}, this
author considered the general class of NLS equations with arbitrary
forms of nonlinearities and potentials in arbitrary spatial
dimensions, and derived the general analytical conditions for
pitchfork bifurcations as well as the power formulae for
solitary-wave branches near the pitchfork bifurcation point.

In this paper, we consider the general nonlinear Schr\"odinger
equations with arbitrary forms of nonlinearity and potentials in
arbitrary spatial dimensions (as in \cite{Yang_classification}).
These equations include the Gross-Pitaevskii equations in
Bose-Einstein condensates with attractive or repulsive atomic
interactions and nonlinear light-transmission equations in linear
potentials or nonlinear lattices with power or non-power
nonlinearities as special cases
\cite{Kivshar_book,Yang_SIAM,BEC,Malomed_nonlinear_lattice}. For
this large class of equations, we determine the linear stability of
both sign-definite (positive) and sign-indefinite solitary waves
near pitchfork bifurcations. Our strategy is to explicitly calculate
the bifurcation of linear-stability eigenvalues from the origin,
which always takes place whenever a pitchfork bifurcation occurs
(see Theorem 3 in Sec. 3). Based on this eigenvalue bifurcation and
assuming no other instabilities interfere, linear stability of
solitary waves near pitchfork bifurcations is then obtained (see
Theorem 4 in Sec. 3). We show that the smooth solution branch always
switches stability at the bifurcation point. In addition, the
bifurcated solution branches and the smooth branch have opposite
(same) stability when their power slopes have the same (opposite)
sign. One unusual feature on the linear stability of these pitchfork
bifurcations is that the smooth and bifurcated solution branches (on
the same side of the bifurcation point) can be both stable or both
unstable, which contrasts such bifurcations in finite-dimensional
dynamical systems where the smooth and bifurcated branches generally
have opposite stability \cite{GH}. For the special case of positive
solitary waves, stronger and more explicit stability results are
also obtained (see Theorem 5 in Sec. 3). We show that for positive
solitary waves, their linear stability near a pitchfork bifurcation
point can be read off directly from their power diagram.
Specifically, their linear stability is simply determined by which
side of the bifurcation point the bifurcated solutions appear and
whose power slope of the smooth and bifurcated solutions is larger.
Lastly, we present various numerical examples, and show that the
numerical results confirm the analytical predictions both
qualitatively and quantitatively.

Compared with the earlier analytical results on stability of
pitchfork bifurcations (such as in
\cite{Weinstein_2004,Weinstein_2008,Sacchetti_2009,Kirr_2011}), our
stability results have the following three distinctive features.
First, our results apply to the general NLS equations with no
restriction on the nonlinearity, potential or spatial dimensions.
Second, we made a direct link between linear stability and the power
diagram (especially for positive solitary waves whose linear
stability can be read off entirely from the power diagram). Third,
we derived explicit analytical formulae for linear-stability
eigenvalues of solitary waves, which can be useful when quantitative
prediction of linear instability is needed.

\section{Preliminaries}
We consider the generalized nonlinear Schr\"odinger (GNLS) equations
with arbitrary forms of nonlinearity and external potentials in
arbitrary spatial dimensions. These equations can be written as
\begin{equation}  \label{e:U}
iU_t+\nabla^2 U+F(|U|^2, \x)  \hspace{0.06cm} U=0,
\end{equation}
where $\nabla^2=\partial^2/\partial x_1^2+\partial^2/\partial
x_2^2+\cdots + \partial^2/\partial x_N^2$ is the Laplacian in the
$N$-dimensional space $\textbf{x}=(x_1, x_2, \cdots, x_N)$, and
$F(\cdot, \cdot)$ is a general real-valued function which includes
nonlinearity as well as external potentials. These GNLS equations
include the Gross-Pitaevskii equations in Bose-Einstein condensates
\cite{BEC} and nonlinear light-transmission equations in linear
potentials and nonlinear lattices
\cite{Kivshar_book,Yang_SIAM,Malomed_nonlinear_lattice} as special
cases. Notice that these equations are conservative and Hamiltonian.

For a large class of nonlinearities and potentials, this equation
admits stationary solitary waves
\begin{equation}  \label{e:Usoliton}
U(\x,t)=e^{i\mu t}u(\x),
\end{equation}
where $u(\x)$ is a real and localized function in the
square-integrable functional space which satisfies the equation
\begin{equation}  \label{e:u}
\nabla^2u-\mu u+F(u^2, \x)  \hspace{0.05cm} u=0,
\end{equation}
and $\mu$ is a real-valued propagation constant. Examples of such
solitary waves can be found in numerous books and articles (see
\cite{Kivshar_book,Yang_SIAM} for instance). In these solitary
waves, $\mu$ is a free parameter, and $u(\x)$ depends continuously
on $\mu$. Under certain conditions, these solitary waves undergo
bifurcations at special values of $\mu$. Three major types of
bifurcations have been classified \cite{Yang_classification}. Of
these bifurcations, stability of solitary waves near saddle-node
bifurcations has been analyzed in \cite{Yang_saddle1,Yang_saddle2}.
It was shown that no stability switching takes place at a
saddle-node bifurcation, which dispels a pervasive misconception
that such stability switching should occur. In this paper, we study
the stability of solitary waves near pitchfork bifurcations.

A pitchfork bifurcation in Eq. (\ref{e:U}) is where on one side of
the bifurcation point $\mu=\mu_0$, there is a single solitary wave
branch $u^0(\x; \mu)$; but on the other side of $\mu_0$, three
distinct solitary-wave branches appear. One of them is a smooth
continuation of the $u^0(\x; \mu)$ branch, but the other two
branches $u^\pm(\x; \mu)$ are new and they bifurcate out at
$\mu=\mu_0$.

To present conditions for pitchfork bifurcations, we introduce the
linearization operator of Eq. (\ref{e:u}),
\begin{equation}  \label{e:L1}
L_1=\nabla^2-\mu+\partial_u[F(u^2, \x)u],
\end{equation}
which is a self-adjoint linear Schr\"odinger operator. We also
introduce the standard inner product of functions,
\begin{equation} \label{def:inner_prod}
\langle f, g\rangle  = \int_{-\infty}^\infty f^*(\x) \hspace{0.05cm}
g(\x) \hspace{0.07cm} d \x,
\end{equation}
where the superscript `*' represents complex conjugation. In
addition, we define the power of a solitary wave $u(\x;\mu)$ as
\begin{equation}
P(\mu)=\langle u, u\rangle = \int_{-\infty}^\infty u^2(\x; \mu)
\hspace{0.07cm} d \x,
\end{equation}
and denote the power functions of the smooth and bifurcated solution
branches as
\begin{equation}
P_0(\mu)\equiv \langle u^0(\x; \mu), u^0(\x; \mu) \rangle, \quad
P_\pm(\mu)\equiv \langle u^\pm(\x; \mu), u^\pm(\x; \mu) \rangle.
\end{equation}

If a bifurcation occurs at $\mu=\mu_0$, by denoting the
corresponding solitary wave and the $L_1$ operator as
\begin{equation}
u_0(\x) \equiv u(\x; \mu_0), \quad L_{10}  \equiv  L_1|_{\mu=\mu_0,\
u=u_0},
\end{equation}
then $L_{10}$ should have a discrete zero eigenvalue. This is a
necessary condition for all bifurcations, not just for pitchfork
bifurcations. In \cite{Yang_classification}, the following
sufficient conditions for pitchfork bifurcations were derived.

\vspace{0.2cm} \textbf{Theorem 1} \ Assume that zero is a simple
discrete eigenvalue of $L_{10}$. Denote the real eigenfunction of
this zero eigenvalue as $\psi(\x)$, and denote
\begin{equation} \label{n:G}
G(u;\x)  \equiv F(u^2;\x) \hspace{0.04cm} u, \quad G_k(\x) \equiv
\partial_u^k G|_{u=u_0}, \ k=2, 3.
\end{equation}
Then if
\begin{equation}
\langle u_0, \psi\rangle = \langle G_2, \psi^3\rangle = 0,
\end{equation}
\begin{equation} \label{def:R}
R\equiv \langle 1-G_2L_{10}^{-1}u_0, \psi^2\rangle\ne 0,
\end{equation}
and
\begin{equation}
S\equiv \langle G_3, \psi^4\rangle- 3\langle G_2\psi^2,
L_{10}^{-1}(G_2\psi^2)\rangle\ne 0,
\end{equation}
a pitchfork bifurcation occurs at $\mu=\mu_0$. The new solitary
waves $u^\pm(\x; \mu)$ bifurcate to the right (left) side of
$\mu=\mu_0$ if the constants $R$ and $S$ have the same (opposite)
sign.

In addition, slopes of power functions for the smooth and bifurcated
solitary-wave branches at the pitchfork bifurcation point were also
derived in \cite{Yang_classification}.

\vspace{0.2cm} \textbf{Theorem 2} \ Suppose the conditions in
Theorem 1 hold and a pitchfork bifurcation occurs at $\mu=\mu_0$.
Then power slopes of the smooth and bifurcated solitary-wave
branches at the bifurcation point are given as
\begin{equation}  \label{f:P0case2}
P_0'(\mu_0)=2 \hspace{0.03cm} \langle u_0, L_{10}^{-1}u_0 \rangle,
\end{equation}
and
\begin{equation} \label{f:P1case2}
P_+'(\mu_0)=P_-'(\mu_0)= P_0'(\mu_0) +\frac{6R^2}{S}.
\end{equation}
Here (and in later text) the prime represents the derivative.

In this article, we consider pitchfork bifurcations in the GNLS
equations (\ref{e:U}) where the bifurcation conditions in Theorem 1
hold.

The main goal of this paper is to determine the linear stability of
solitary waves near these pitchfork bifurcations. To study this
linear stability, we perturb the solitary waves as \cite{Yang_SIAM}
\begin{equation}\label{e:perturb}
U(\textbf{x},t)=e^{i\mu t} \left\{ {u (\textbf{x}
)+[{v}(\textbf{x})+w(\textbf{x})]e^{\lambda
t}+[{v}^*(\textbf{x})-w^* (\textbf{x})]e^{\lambda^* t} }\right\} ,
\end{equation}
where $v, w\ll 1$ are normal-mode perturbations, and $\lambda$ is
the mode's eigenvalue. Inserting this perturbed solution into
(\ref{e:U}) and linearizing, we obtain the following linear
eigenvalue problem
\begin{equation} \label{e:LPhi}
{\cal L}\Phi=-i\lambda \Phi,
\end{equation}
where
\begin{equation} \label{e:calL}
{\cal L} =  \left[\begin{array}{cc} 0 & L_0 \\ L_1 & 0
\end{array}\right], \quad \Phi = \left[\begin{array}{c} v \\ w
\end{array}\right],
\end{equation}
\begin{equation} \label{e:L0}
 L_0= \nabla^2-\mu+F(u^2, \x),
\end{equation}
and $L_1$ is as defined in Eq. (\ref{e:L1}). Both $L_0$ and $L_1$
are linear Schr\"odinger operators and are Hermitian. In the later
text, operator ${\cal L}$ will be called the linear-stability
operator. The eigenvalue problem (\ref{e:LPhi}) can also be written
as
\begin{equation}  \label{e:L0wL1v}
L_0w=-i\lambda v, \quad L_1v=-i\lambda w.
\end{equation}
If this linear-stability eigenvalue problem admits eigenvalues
$\lambda$ whose real parts are positive, then the corresponding
normal-mode perturbation in Eq. (\ref{e:perturb}) exponentially
grows, hence the solitary wave $u(\x)$ is linearly unstable.
Otherwise it is linearly stable. Notice that eigenvalues of this
linear-stability problem always appear in quadruples $(\lambda,
-\lambda, \lambda^*, -\lambda^*)$ when $\lambda$ is complex, or in
pairs $(\lambda, -\lambda)$ when $\lambda$ is real or purely
imaginary.

Using the $L_0$ operator, the solitary wave equation (\ref{e:u}) can
be written as
\begin{equation}  \label{e:L0u}
L_0u=0.
\end{equation}
Differentiating this equation with respect to $\mu$, we find that
\begin{equation}  \label{e:L1umu}
L_1u_\mu=u,
\end{equation}
where $u_\mu\equiv \partial u/\partial \mu$. These two relations
will be useful in later analysis. Due to (\ref{e:L0u}), the
linear-stability eigenvalue problem (\ref{e:L0wL1v}) admits a zero
eigenmode for every solitary wave $u(\x; \mu)$:
\begin{equation} \label{e:zeromode}
\lambda=0, \quad v=0, \quad w=u.
\end{equation}
This zero eigenmode is related to the phase invariance of the GNLS
equation (\ref{e:U}), which says that if $U(\x,t)$ is a solution of
(\ref{e:U}), so is $U(\x,t) \hspace{0.05cm} e^{i\theta}$ for any
real phase constant $\theta$.

The GNLS equations (\ref{e:U}) may be viewed as an
infinite-dimensional dynamical system, with solitary waves
(\ref{e:Usoliton}) being its fixed points. In this view, it is
tempting to deduce the stability of pitchfork bifurcations in the
GNLS equations (\ref{e:U}) from those in finite-dimensional
dynamical systems. In finite-dimensional dynamical systems, it has
been shown that at a pitchfork bifurcation point, the smooth
fixed-point branch changes its stability. In addition, the two
bifurcated fixed-point branches have the opposite stability of the
smooth fixed-point branch (on the same side of the bifurcation
point) \cite{GH}. However, these stability results were derived
under the assumption that zero is a simple eigenvalue of the
Jacobian (linearization) matrix of the system at the bifurcation
point (see Ref. \cite{GH}, Theorem 3.4.1, Hypothesis SN1). For the
GNLS equations (\ref{e:U}), the counterpart of the Jacobian matrix
is the linear-stability operator ${\cal L}$ defined in Eq.
(\ref{e:calL}), but zero is \emph{not} a simple eigenvalue of ${\cal
L}$ at the bifurcation point (see Eq. (\ref{e:L0zeromode}) below).
This means that we cannot apply the above stability results from
finite-dimensional dynamical systems to pitchfork bifurcations in
the GNLS equations (\ref{e:U}). Instead we have to analyze this
stability for Eq. (\ref{e:U}) separately. As we will see, stability
for pitchfork bifurcations in Eq. (\ref{e:U}) shows some novel
features which do not exist in finite-dimensional dynamical systems.
It is relevant to mention that the same situation arises for
saddle-node bifurcations as well, where it was shown in
\cite{Yang_saddle1,Yang_saddle2} that no stability switching occurs
in the GNLS equations (\ref{e:U}) even though such stability
switching generally takes place in finite-dimensional dynamical
systems \cite{GH}.

\section{Main results}
Our stability analysis starts with the basic fact that, at a
pitchfork bifurcation point $\mu=\mu_0$, $L_{10}$ has a discrete
zero eigenvalue (see earlier text). With the eigenfunction of this
zero eigenvalue denoted as $\psi(\x)$ (see Theorem 1), we have
\begin{equation} \label{e:L10psi}
L_{10}\psi=0.
\end{equation}
Thus at the bifurcation point $\mu=\mu_0$, in addition to the
phase-invariance-induced zero eigenmode (\ref{e:zeromode}), the
linear-stability eigenvalue problem (\ref{e:L0wL1v}) also admits
another bifurcation-induced zero eigenmode,
\begin{equation} \label{e:zeromode2}
\lambda=0, \quad v=\psi, \quad w=0.
\end{equation}
Away from the bifurcation point ($\mu\ne \mu_0$), while the
phase-related zero eigenvalue (\ref{e:zeromode}) persists at the
origin, the bifurcation-induced zero eigenvalue (\ref{e:zeromode2})
bifurcates out since this zero eigenmode does not exist in $L_{1}$
any more. Thus our approach is to determine how this
bifurcation-induced zero eigenvalue moves out of the origin when
$\mu$ moves away from $\mu_0$. We will show that this zero
eigenvalue only bifurcates out along the real or imaginary axis as a
$\pm \lambda$ pair. Bifurcation along the real axis creates
instability, while bifurcation along the imaginary axis does not
create instability. Thus, based on which direction this zero
eigenvalue bifurcates and assuming no other instabilities interfere,
linear-stability behaviors of solitary waves near the bifurcation
point will be analytically obtained. In the special case of positive
solitary waves, we will show that there is indeed no other
instabilities interfering near a pitchfork bifurcation, hence
stronger and more explicit stability results will be derived.

For later analysis, we introduce two additional notations,
\begin{equation} \label{e:L00L0}
L_{00} \equiv L_0|_{\mu=\mu_0,\ u=u_0}, \quad  {\cal L}_0 \equiv
{\cal L}_{\mu=\mu_0,\ u=u_0}.
\end{equation}
In view of Eq. (\ref{e:L0u}), we have
\begin{equation}  \label{e:L00u0}
L_{00}u_0=0,
\end{equation}
thus zero is a discrete eigenvalue of $L_{00}$. From Eqs.
(\ref{e:zeromode}) and (\ref{e:zeromode2}), we have
\begin{equation} \label{e:L0zeromode}
{\cal L}_0 \left[\begin{array}{c} 0 \\ u_0 \end{array}\right]={\cal
L}_0 \left[\begin{array}{c} \psi \\ 0 \end{array}\right]=0,
\end{equation}
so zero is also a (multifold) discrete eigenvalue of ${\cal L}_0$.

On the bifurcation of the zero eigenvalue in the linear-stability
operator ${\cal L}$ when $\mu \ne \mu_0$, we have the following main
results.

\vspace{0.3cm} \textbf{Theorem 3} \ Assume that zero is a simple
discrete eigenvalue of $L_{00}$ and $L_{10}$. Near a pitchfork
bifurcation point $\mu=\mu_0$ in Theorem 1, if
\begin{equation}  \label{c:pitchfork2}
\langle \psi, L_{00}^{-1}\psi\rangle \ne 0, \quad P_0'(\mu_0) \ne 0,
\quad P_\pm'(\mu_0) \ne 0,
\end{equation}
then a single pair of non-zero eigenvalues $\pm \lambda$ in ${\cal
L}$ bifurcate out along the real or imaginary axis from the origin
when $\mu\ne \mu_0$;
\renewcommand{\labelenumi}{(\alph{enumi})}
\begin{enumerate}
\item
on the smooth solution branch $u^0(\x; \mu)$, the bifurcated
eigenvalues $\lambda^0$ are given asymptotically by
\begin{equation} \label{f:lambdacase2a}
(\lambda^0)^2 \to \alpha \hspace{0.06cm} (\mu-\mu_0), \quad \mu \to
\mu_0,
\end{equation}
where the real constant $\alpha$ is
\begin{equation}  \label{f:alpha2}
\alpha=\frac{R}{\langle \psi, L_{00}^{-1}\psi\rangle} \ne 0;
\end{equation}

\item
on the two bifurcated solution branches $u^\pm(\x; \mu)$, the
bifurcated eigenvalues $\lambda^\pm$ are given asymptotically by
\begin{equation} \label{f:lambdapmcase2b}
(\lambda^\pm)^2 \to \beta \hspace{0.04cm} (\mu-\mu_0), \quad \mu \to
\mu_0,
\end{equation}
where the real constant $\beta$ is
\begin{equation} \label{f:beta}
\beta = - 2\alpha \frac{P_\pm'(\mu_0)}{P_0'(\mu_0)}  \ne 0.
\end{equation}
\end{enumerate}

\vspace{0.1cm} \textbf{Remark 1} \ Due to the assumption in this
theorem, the discrete zero eigenvalue in ${\cal L}_0$ is not
embedded inside its continuous spectrum. This fact allows us to
calculate eigenvalue bifurcations from the origin in $\cal L$ when
$0<|\mu-\mu_0|\ll 1$ by perturbation series expansions (without
worrying about continuous-wave tails in the eigenfunctions beyond
all orders of the perturbation expansion
\cite{Yang_SIAM,Pomeau_1988,Grimshaw_1995,Calvo_Akylas_1997}).

\textbf{Remark 2} \ In this theorem, the assumption of zero being a
simple discrete eigenvalue of $L_{00}$ and $L_{10}$ is satisfied in
all one-dimensional bifurcations and many higher-dimensional
bifurcations.

A direct consequence of Theorem 3 is the following Theorem 4 which
summarizes the qualitative linear-stability properties of solitary
waves near a pitchfork bifurcation point.

\vspace{0.2cm} \textbf{Theorem 4} \ Suppose at a pitchfork
bifurcation point $\mu=\mu_0$, the solitary wave $u_0(\x)$ is
linearly stable (i.e., all its eigenvalues are either zero or purely
imaginary); and when $\mu$ moves away from $\mu_0$, no complex
eigenvalues bifurcate out from non-zero points on the imaginary
axis. Then under the same conditions of Theorem 3, the smooth
solution branch $u^0(\x; \mu)$ undergoes stability switching at the
bifurcation point [with the right (left) side being unstable if the
constant $\alpha$ in (\ref{f:alpha2}) is positive (negative)]. Near
the bifurcation point, the two bifurcated solution branches
$u^\pm(\x; \mu)$ and the smooth solution branch (on the same side of
the bifurcation point) have opposite (same) linear stability when
their power slopes $P_0'(\mu_0)$ and $P_\pm'(\mu_0)$ have the same
(opposite) sign.

\begin{figure}[h!]
\centerline{\includegraphics[width=0.8\textwidth]{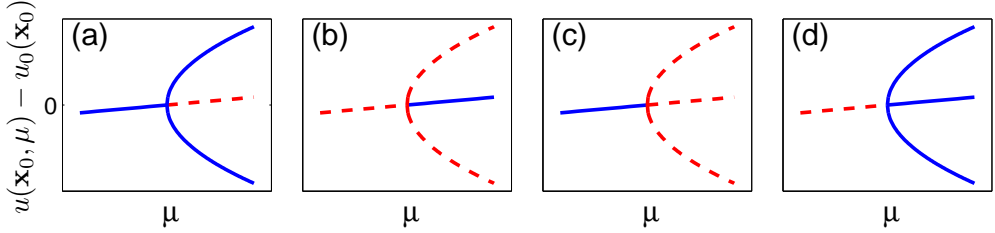}}
\caption{Schematic diagrams for the four types of pitchfork
bifurcations in the GNLS equations (\ref{e:U}). Plotted in this
figure are deviations $u(\x_0; \mu)-u_0(\x_0)$ versus $\mu$ at a
representative $\x_0$ position. Solid blue and dashed red lines
indicate stable and unstable branches respectively. } \label{f:fig1}
\end{figure}

\vspace{0.1cm} Based on this theorem, there are four types of
pitchfork bifurcations in the GNLS equations (\ref{e:U}), and their
schematic diagrams are displayed in Fig.~\ref{f:fig1} (note that
horizontal flips of these bifurcation diagrams are also admissible,
but we will not distinguish them from the ones in Fig.~\ref{f:fig1}
for brevity). The bifurcations in Fig.~\ref{f:fig1}(a,b) are
qualitatively the same as the supercritical and subcritical
pitchfork bifurcations in finite-dimensional dynamical systems
\cite{GH}. In these two cases, the bifurcated solution branches and
the smooth solution branch have opposite stability. The bifurcations
in Fig.~\ref{f:fig1}(c,d), however, are different. In these two
cases, the bifurcated solution branches and the smooth branch have
the \emph{same} stability (all stable or all unstable), which seems
to have no counterpart in the classical dynamical-system
theory~\cite{GH}. Note that the bifurcation in Fig.~\ref{f:fig1}(c)
has been reported in \cite{Kirr_2011}, but the bifurcation in
Fig.~\ref{f:fig1}(d) has not been discovered before to the author's
best knowledge.

\textbf{Remark 3} \ In Theorem 4, the sign of $\alpha$ plays a
critical role for the stability outcome. This sign can be determined
as follows. From formula (\ref{f:alpha2}), we see that the sign of
$\alpha$ is determined by the signs of $R$ and $\langle \psi,
L_{00}^{-1}\psi\rangle$. We know from Theorem~1 that $R$ and $S$
have the same (opposite) sign if the bifurcated solitary waves
appear on the right (left) side of $\mu=\mu_0$. We also know from
formula (\ref{f:P1case2}) that $S$ and $P_\pm'(\mu_0)-P_0'(\mu_0)$
have the same sign. Using this information, the sign of $\alpha$ can
be determined as follows.
\renewcommand{\labelenumi}{(\roman{enumi})}
\begin{enumerate}
\item If the bifurcated solitary waves appear on the right side of $\mu=\mu_0$, then the
sign of $\alpha$ is equal to the sign of $\langle \psi,
L_{00}^{-1}\psi\rangle$ multiplying the sign of
$P_\pm'(\mu_0)-P_0'(\mu_0)$;
\item If the bifurcated solitary waves appear on the left side of $\mu=\mu_0$, then the
sign of $\alpha$ is opposite of the sign of $\langle \psi,
L_{00}^{-1}\psi\rangle$ multiplying the sign of
$P_\pm'(\mu_0)-P_0'(\mu_0)$.
\end{enumerate}

\begin{figure}[h!]
\centerline{\includegraphics[width=0.65\textwidth]{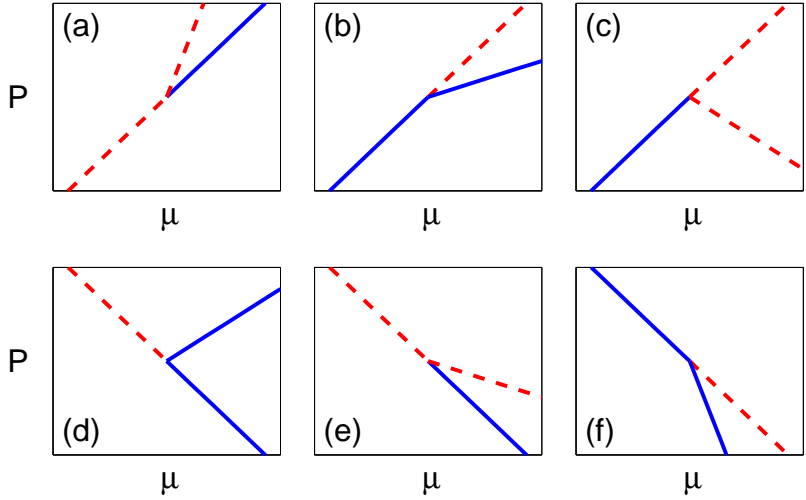}}
\caption{Six possible scenarios of the power diagram (with stability
indicated) for pitchfork bifurcations in the GNLS equations
(\ref{e:U}) when $\langle \psi, L_{00}^{-1}\psi\rangle<0$ and the
bifurcated solutions reside on the right side of the bifurcation
point. Solid blue and dashed red lines indicate stable and unstable
branches respectively. } \label{f:fig2}
\end{figure}

From this remark, we see that if the sign of $\langle \psi,
L_{00}^{-1}\psi\rangle$ is known, then the sign of $\alpha$ can be
read off from the structure of the power diagram (i.e., from which
side of $\mu=\mu_0$ the bifurcated solutions reside and which of the
power slopes $P_\pm'(\mu_0)$ and $P_0'(\mu_0)$ is larger). After the
sign of $\alpha$ is obtained, stability of all the solution branches
can then be read off again from the power diagram by Theorem 4. For
instance, when $\langle \psi, L_{00}^{-1}\psi\rangle<0$ and the
bifurcated solutions appear on the right side of the bifurcation
point $\mu=\mu_0$, schematic power diagrams of all six possible
bifurcation scenarios (with stability information indicated) are
displayed in Fig. \ref{f:fig2}. This list of power diagrams is
compiled according to the signs of $P_0'(\mu_0)$, $P_\pm'(\mu_0)$,
and $P_\pm'(\mu_0)-P_0'(\mu_0)$. If $\langle \psi,
L_{00}^{-1}\psi\rangle>0$, these power diagrams remain the same, but
the stability of all branches is flipped (with ``stable" changed to
``unstable" and vise versa). If the bifurcated solutions appear on
the left side of the bifurcation point $\mu=\mu_0$, their power
diagrams (with stability information) can be similarly obtained.

One might notice that the power diagrams of pitchfork bifurcations
in Fig.~\ref{f:fig2} split out to two rather than three branches at
the bifurcation point, which is different from the
solution-bifurcation diagrams in Fig.~\ref{f:fig1}. The reason is
that power slopes of the two bifurcated solution branches $u^\pm(\x;
\mu)$ are the same at the bifurcation point (see Theorem 2), thus
one sees only two power branches instead of three
\cite{Yang_classification}.

The above results (i.e., Theorems 1--4) are valid for all
real-valued solitary waves $u(\x; \mu)$ in the GNLS equations
(\ref{e:U}), including both sign-definite (positive) and
sign-indefinite (sign-changing) solitary waves. If the solitary
waves are positive, then our stability results can be made stronger
and more explicit. For positive solitary waves in Eq. (\ref{e:U}),
it is known that all eigenvalues in the linear-stability operator
${\cal L}$ are either purely real or purely imaginary (see Ref.
\cite{Yang_SIAM}, Theorem 5.2). In addition, linear stability of the
solitary wave $u_0(\x)$ at the bifurcation point can be determined
by the generalized Vakhitov-Kolokolov stability criterion
\cite{Yang_SIAM}. Furthermore, zero is the largest eigenvalue of
$L_0$ and is simple \cite{Struwe}, and $\langle \psi,
L_{00}^{-1}\psi\rangle<0$ since operator $L_{00}$ is semi-negative
definite. Using this information, together with Theorem 4 and Remark
3, we can obtain the following stronger and more explicit theorem
for the linear stability of positive solitary waves near a pitchfork
bifurcation point. This theorem derives linear stability of these
solitary waves almost exclusively from their power diagram.

\vspace{0.2cm} \textbf{Theorem 5} \ Suppose solitary waves in the
GNLS equations (\ref{e:U}) are positive near a pitchfork bifurcation
point $\mu=\mu_0$. If $P_0'(\mu_0)<0$, or zero is the $n$-th largest
discrete eigenvalue of $L_{10}$ with $n\ge 3$, then these solitary
waves (near the bifurcation point) are linearly unstable. If
$P_0'(\mu_0)>0$, $P_\pm'(\mu_0)\ne 0$, and zero is the second
largest and simple discrete eigenvalue of $L_{10}$, then
\renewcommand{\labelenumi}{(\arabic{enumi})}
\begin{enumerate}
\item when the bifurcated solitary-wave branches $u^\pm(\x; \mu)$ appear on the right side of
$\mu=\mu_0$ and $P_\pm'(\mu_0) < P_0'(\mu_0)$, the smooth solution
branch $u^0(\x; \mu)$ is linearly stable for $\mu<\mu_0$ and
unstable for $\mu>\mu_0$, whereas the bifurcated branches are
linearly stable if $P_\pm'(\mu_0) >0$ and unstable if $P_\pm'(\mu_0)
<0$;

\item when the bifurcated solitary waves appear on the right side of
$\mu=\mu_0$ and $P_\pm'(\mu_0) > P_0'(\mu_0)$, the smooth solution
branch is linearly unstable for $\mu<\mu_0$ and stable for
$\mu>\mu_0$, whereas the bifurcated branches are always linearly
unstable;

\item when the bifurcated solitary waves appear on the left side of
$\mu=\mu_0$ and $P_\pm'(\mu_0) < P_0'(\mu_0)$, the smooth solution
branch is linearly unstable for $\mu<\mu_0$ and stable for
$\mu>\mu_0$, whereas the bifurcated branches are linearly stable if
$P_\pm'(\mu_0) >0$ and unstable if $P_\pm'(\mu_0) <0$;

\item when the bifurcated solitary waves appear on the left side of
$\mu=\mu_0$ and $P_\pm'(\mu_0) > P_0'(\mu_0)$, the smooth solution
branch is linearly stable for $\mu<\mu_0$ and unstable for
$\mu>\mu_0$, whereas the bifurcated branches are always linearly
unstable.
\end{enumerate}

In terms of solution-bifurcation diagrams, case (1) in this theorem
belongs to pitchfork bifurcations of type (a) or (c) in Fig. 1, case
(2) belongs to pitchfork bifurcations of type (b) in Fig. 1, case
(3) belongs to pitchfork bifurcations of type (a) or (c) in Fig. 1
with the horizontal axis flipped, and case (4) belongs to pitchfork
bifurcations of type (b) in Fig. 1 with the horizontal axis flipped.
Thus for positive solitary waves in the GNLS equations (\ref{e:U}),
pitchfork bifurcations of type (d) in Fig. 1 (and its horizontal
flip) cannot occur.

In terms of power-bifurcation diagrams, case (1) belongs to
pitchfork bifurcations of type Fig. 2(b,c), case (2) belongs to
pitchfork bifurcations of type Fig. 2(a), case (3) belongs to
pitchfork bifurcations of type Fig. 2(d,e) with horizontal-axis
flipping as well as branch-stability flipping, and case (4) belongs
to pitchfork bifurcations of type Fig. 2(f) with horizontal-axis
flipping as well as branch-stability flipping.

\textbf{Remark 4} \ For positive solitary waves, pitchfork
bifurcation cannot occur when the largest discrete eigenvalue of
$L_{10}$ is zero. The reason is that for any linear Schr\"odinger
operator, the eigenfunction of its largest eigenvalue is always
positive (sign-definite) \cite{Struwe}. Thus for this largest zero
eigenvalue of $L_{10}$, its eigenfunction $\psi$ is positive. Since
$u_0(\x)$ is also positive, then $\langle u_0, \psi\rangle \ne 0$,
which violates the conditions of pitchfork bifurcations in Theorem
1. Thus this case is not mentioned in Theorem~5.

\section{Proofs of the main results}

\textbf{Proof of Theorem 3} \ The basic idea of the proof is that we
first show the algebraic multiplicity of the zero eigenvalue in the
linear-stability operator $\cal L$ is four at the bifurcation point
$\mu=\mu_0$ and drops to two away from it, thus a pair of
eigenvalues bifurcate out from the origin when $\mu\ne \mu_0$. This
pair of eigenvalues must bifurcate along the real or imaginary axis
since eigenvalues of $\cal L$ would appear as quadruples if this
bifurcation were not along these two axes. Then we calculate this
pair of eigenvalues near the bifurcation point $\mu=\mu_0$ by
perturbation methods. We show that the perturbation series for these
bifurcated eigenvalues can be constructed to all orders, with the
leading-order terms given by Eqs. (\ref{f:lambdacase2a}) and
(\ref{f:lambdapmcase2b}) for solitary waves on the smooth and
bifurcated branches respectively.

At the pitchfork bifurcation point $\mu=\mu_0$ in Theorem 1, $(0,
u_0)^T$ and $(\psi, 0)^T$ are eigenfunctions of the zero eigenvalue
in ${\cal L}_0$ in view of Eq. (\ref{e:L0zeromode}). Here the
superscript `$T$' represents the transpose of a vector.  Under the
assumption in Theorem 3, ${\cal L}_0$ does not admit any additional
eigenfunctions at the zero eigenvalue, thus the geometric
multiplicity of this zero eigenvalue is two. Next we determine the
algebraic multiplicity of this zero eigenvalue by examining its
generalized eigenfunctions.

First, evaluating the relation (\ref{e:L1umu}) along the smooth
solution branch $u^0(\x; \mu)$ at $\mu=\mu_0$, we get
\begin{equation}
L_{10} \hspace{0.02cm} u^0_{\mu 0}=u_0,
\end{equation}
where $u^0_{\mu 0}$ is equal to $u^0_{\mu}$ evaluated at
$\mu=\mu_0$. Thus,
\begin{equation}
{\cal L}_0\left[\begin{array}{c} u^0_{\mu 0} \\ 0
\end{array}\right]=\left[\begin{array}{c} 0 \\ u_0
\end{array}\right],
\end{equation}
which means that $(u^0_{\mu 0}, 0)^T$ is a generalized eigenfunction
of the zero eigenvalue. The next-order generalized eigenfunction
$(f_2, g_2)^T$ satisfies the equation
\begin{equation}
{\cal L}_0 \left[\begin{array}{c} f_2 \\ g_2
\end{array}\right]=\left[\begin{array}{c} u^0_{\mu 0} \\ 0
\end{array}\right],
\end{equation}
so the equation for $g_2$ is
\begin{equation} \label{e:L00g2umu}
L_{00}g_2=u^0_{\mu 0}.
\end{equation}
Since $u_0$ is a homogeneous solution of this equation, $L_{00}$ is
self-adjoint, and $\langle u_0, u^0_{\mu 0}\rangle=P_0'(\mu_0)/2\ne
0$ by conditions (\ref{c:pitchfork2}), according to the Fredholm
alternative theorem, Eq. (\ref{e:L00g2umu}) does not admit localized
solutions for $g_2$, thus there are no additional generalized
eigenfunctions for $(0, u_0)^T$.

Next, we consider generalized eigenfunctions of ${\cal L}_0$ for
$(\psi, 0)^T$. The lowest-order generalized eigenfunction $(f_1,
g_1)$ satisfies the equation
\begin{equation}  \label{e:L0f1g1psi}
{\cal L}_0 \left[\begin{array}{c} f_1 \\ g_1
\end{array}\right]=\left[\begin{array}{c} \psi \\ 0
\end{array}\right].
\end{equation}
According to the assumption in Theorem 3, the kernel of $L_{00}$
contains a single localized function $u_0$, and $\langle u_0,
\psi\rangle=0$ in view of the conditions for pitchfork bifurcations
in Theorem 1. Thus from the Fredholm alternative theorem, there
exists a real localized function $L_{00}^{-1}\psi$. Consequently,
\begin{equation}  \label{e:L0generalizedcase2}
{\cal L}_0 \left[\begin{array}{c} 0 \\ L_{00}^{-1}\psi
\end{array}\right]=\left[\begin{array}{c} \psi \\ 0
\end{array}\right],
\end{equation}
i.e., $(0, L_{00}^{-1}\psi)^T$ is a generalized eigenfunction of
${\cal L}_0$. The next-order generalized eigenfunction $(f_2,
g_2)^T$ for $(\psi, 0)^T$ satisfies the equation
\begin{equation}
{\cal L}_0 \left[\begin{array}{c} f_2 \\ g_2
\end{array}\right]=\left[\begin{array}{c} 0 \\ L_{00}^{-1}\psi
\end{array}\right],
\end{equation}
so the equation for $f_2$ is
\begin{equation} \label{e:L10f2L0psi}
L_{10}f_2=L_{00}^{-1}\psi.
\end{equation}
Since $\psi$ is a homogeneous solution of this equation and $\langle
\psi, L_{00}^{-1}\psi\rangle\ne 0$ by conditions
(\ref{c:pitchfork2}), Eq. (\ref{e:L10f2L0psi}) does not admit any
localized solutions by the Fredholm alternative theorem. Thus there
are no additional generalized eigenfunctions for $(\psi, 0)^T$.

The above analysis shows that ${\cal L}_0$ has two eigenfunctions
and two generalized eigenfunctions at the zero eigenvalue, thus the
algebraic multiplicity of the zero eigenvalue in ${\cal L}_0$ (at
the bifurcation point) is four.

When $\mu\ne \mu_0$, on any solitary-wave branch $u^0(\x; \mu)$ or
$u^\pm(\x; \mu)$, Eqs. (\ref{e:L0u})-(\ref{e:L1umu}) hold, thus $(0,
u)^T$ is an eigenfunction of the zero eigenvalue in ${\cal L}$ (see
Eq. (\ref{e:zeromode})), and $(u_\mu, 0)^T$ is its generalized
eigenfunction. This eigenfunction and generalized eigenfunction at
$\mu\ne \mu_0$ are the counterparts of $(0, u_0)^T$ and $(u^0_{\mu
0}, 0)^T$ at $\mu=\mu_0$ (see above), and they are induced by the
phase invariance of Eq. (\ref{e:U}). Under conditions in Theorem 3,
we can further show, by similar analysis as above, that when
$0<|\mu-\mu_0|\ll 1$, ${\cal L}$ does not admit any additional
eigenfunctions or generalized eigenfunctions at the zero eigenvalue,
thus the algebraic multiplicity of this zero eigenvalue in $\cal L$
is two.

Since the algebraic multiplicity of the zero eigenvalue in $\cal L$
is four at $\mu=\mu_0$ and drops to two when $\mu\ne \mu_0$, this
means that when $\mu$ moves away from $\mu_0$, a pair of
linear-stability eigenvalues must bifurcate out from the origin.
Notice that the two multiplicities of the zero eigenvalue associated
with the phase-invariance mode (\ref{e:zeromode}) persist when
$\mu\ne \mu_0$ (see above), it is then clear that the two non-zero
eigenvalues must bifurcate out from the bifurcation-induced zero
eigenmode (\ref{e:zeromode2}). Since eigenvalues of the
linear-stability operator $\cal L$ always appear in quadruples (when
they are complex) or in pairs (when they are real or purely
imaginary) (see discussions below Eq. (\ref{e:L0wL1v})), this
bifurcated pair of eigenvalues then must be real or purely imaginary
and be opposite of each other as a $\pm \lambda$ pair.

Next, we calculate this pair of bifurcated eigenvalues on the
solution branches $u^0(\x; \mu)$ and $u^\pm(\x; \mu)$. Since at the
bifurcation point the zero eigenvalue of ${\cal L}_0$ is not
embedded inside ${\cal L}_0$'s continuous spectrum, this allows us
to calculate this eigenvalue bifurcation by the perturbation methods
(see Remark 1).

\textbf{(a) Eigenvalue bifurcation along the smooth solution branch}

We first calculate this eigenvalue bifurcation along the smooth
solution branch $u^0(\x; \mu)$. These solitary waves near the
bifurcation point $\mu=\mu_0$ have the following perturbation series
expansion
\begin{equation} \label{e:uexpandcase2a}
u^0(\x; \mu) = \sum_{k=0}^\infty (\mu-\mu_0)^{k}u_k(\x),
\end{equation}
where $u_1, u_2, \dots$ are all real functions
\cite{Yang_classification}. As a consequence, operators $L_0$ and
$L_1$ on this smooth solution branch can be expanded as
\begin{equation}  \label{e:L0L1expandcase2}
L_0^0=\sum_{k=0}^\infty (\mu-\mu_0)^{k} L_{0k}, \quad
L_1^0=\sum_{k=0}^\infty (\mu-\mu_0)^{k} L_{1k}.
\end{equation}
The linear-stability eigenmodes $(v, w, \lambda)$ bifurcated from
the zero eigenmode (\ref{e:zeromode2}) have the following
perturbation series expansions:
\begin{eqnarray}
&& v^0(\x;\mu)=\sum_{k=0}^\infty (\mu-\mu_0)^{k} v_k(\x),   \label{e:vexpand} \\
&& w^0(\x; \mu)  = \lambda_0 (\mu-\mu_0)^{1/2}\sum_{k=0}^\infty
(\mu-\mu_0)^{k}
w_k(\x),   \label{e:wexpand} \\
&& \lambda^0(\mu) =
i\lambda_0(\mu-\mu_0)^{1/2}\left(1+\sum_{k=1}^\infty (\mu-\mu_0)^{k}
\lambda_k\right).           \label{e:lambdaexpand}
\end{eqnarray}
Below we construct these perturbation series solutions to all
orders, and show that the leading-order expressions for the
bifurcated eigenvalues $\lambda^0(\mu)$ are given by the formula
(\ref{f:lambdacase2a}).

We start by substituting the above perturbation expansions into the
linear-stability eigenvalue problem (\ref{e:L0wL1v}). From these
equations at various orders of $\mu-\mu_0$, we get a sequence of
linear equations for $(v_k, w_k)$:
\begin{eqnarray}
&& \hspace{-0.5cm}
L_{10}v_0=0,    \label{e:L10v0} \\
&& \hspace{-0.5cm}
L_{00}w_0=v_0,   \label{e:L00w0}  \\
&& \hspace{-0.5cm}
L_{10}v_1=\lambda_0^2w_0-L_{11}v_0,   \label{e:L10v1}  \\
&& \hspace{-0.5cm}
L_{00}w_1=v_1+\lambda_1v_0-L_{01}w_0, \label{e:L00w1} \\
&& \hspace{-0.5cm}
L_{10}v_2=\lambda_0^2(w_1+\lambda_1w_0)-(L_{11}v_1+L_{12}v_0),   \label{e:L10v2}  \\
&& \hspace{-0.5cm}
L_{00}w_2=v_2+\lambda_1v_1+\lambda_2v_0-(L_{01}w_1+L_{02}w_0),\label{e:L00w2}
\\
&& \hspace{-0.5cm}
\dots\dots\dots  \nonumber \\
&& \hspace{-0.5cm}
L_{10}v_{n+1}=\lambda_0^2\left(w_n+\sum_{k=1}^n \lambda_k w_{n-k}\right)-\sum_{k=1}^{n+1}L_{1k}v_{n+1-k},  \label{e:L10vnp1} \\
&& \hspace{-0.5cm}
L_{00}w_{n+1}=v_{n+1}+\sum_{k=1}^{n+1}\lambda_kv_{n+1-k}-\sum_{k=1}^{n+1}L_{0k}w_{n+1-k},
\label{e:L00wnp1}
\\
&& \hspace{-0.5cm}
\dots\dots\dots  \nonumber
\end{eqnarray}
All these equations are inhomogeneous except the first equation for
$v_0$. From the assumption in Theorem 3, the $v_n$ equations have a
single homogeneous solution $\psi$, and the $w_n$ equations have a
single homogeneous solution $u_0$. Since operators $L_{00}$ and
$L_{10}$ in these equations are self-adjoint, the Fredholm
alternative theorem says that the inhomogeneous equations above are
solvable if and only if their right hand sides are orthogonal to
their homogeneous solution. These orthogonality conditions, together
with a scaling of the eigenfunction $(v, w)$, will determine the
eigenvalue coefficients $\lambda_n$ as well as functions $(v_n,
w_n)$ for all $n\ge 0$, as will be demonstrated below.

First we consider the $v_0$ equation (\ref{e:L10v0}). In view of the
assumption in Theorem 3, the only solution to this equation (after
eigenfunction scaling) is
\begin{equation}
v_0=\psi.
\end{equation}
For the $w_0$ equation (\ref{e:L00w0}), due to the condition of
$\langle u_0, \psi\rangle=0$ for pitchfork bifurcations in Theorem
1, the Fredholm condition is satisfied, thus this equation admits a
real localized solution $L_{00}^{-1}\psi$, and its general solution
is
\begin{equation} \label{f:w0}
w_0=L_{00}^{-1}\psi + c_0 u_0,
\end{equation}
where $c_0$ is a constant to be determined from the solvability
condition of the $w_1$ equation later.

For the $v_1$ equation (\ref{e:L10v1}), it is solvable if and only
if its right hand side is orthogonal to $\psi$. Utilizing the $v_0$
and $w_0$ solutions derived above, this orthogonality condition
yields the formula for the eigenvalue coefficient $\lambda_0$ as
\begin{equation}  \label{f:lambda0}
\lambda_0^2=\frac{\langle \psi, L_{11}\psi\rangle}{\langle \psi,
L_{00}^{-1}\psi\rangle}.
\end{equation}
According to our conditions (\ref{c:pitchfork2}), the denominator in
the above formula is non-zero, thus $\lambda_0^2$ is well defined
and is real. Later in this proof, we will derive a more explicit
expression for $\lambda_0^2$ and show that it is non-zero as well
[see Eq. (\ref{f:lambda02})]. The above equation (\ref{f:lambda0})
gives two real or purely imaginary $\lambda_0$ values as a `$\pm$'
pair.

With the eigenvalue coefficient $\lambda_0$ given in
(\ref{f:lambda0}), the orthogonality condition of the $v_1$ equation
(\ref{e:L10v1}) is satisfied, thus the general solution for $v_1$ is
\begin{equation} \label{f:v1}
v_1=\widehat{v}_1+\lambda_0^2c_0L_{10}^{-1}u_0+d_1\psi,
\end{equation}
where $\widehat{v}_1$ is a real and localized particular solution to
the $v_1$ equation (\ref{e:L10v1}) but without the $c_0$ term in
$w_0$ on its right hand side (see (\ref{f:w0})). This $c_0$ term
induces its own particular solution in $v_1$, which is the middle
term in (\ref{f:v1}). In this term, $L_{10}^{-1}u_0$ is a real
localized function which exists since $u_0$ is orthogonal to the
function $\psi$ in the kernel of $L_{10}$. The last term in
(\ref{f:v1}) is the homogeneous solution, where $d_1$ is a free real
constant. Since this homogeneous term in $v_1$ can be lumped to the
$v_0$ term as $v_0=[1+d_1(\mu-\mu_0)]\psi$ and then eliminated by a
scaling of the eigenfunction $(v, w)$, we will set $d_1=0$. A
similar treatment will be applied to all higher $v_n$ solutions.

Now we consider the $w_1$ equation (\ref{e:L00w1}). Its solvability
condition is that its right hand side be orthogonal to the
homogeneous solution $u_0$. Noticing that $v_0$ is orthogonal to
$u_0$ and $L_{01}$ is a real function, this solvability condition
then reduces to
\begin{equation} \label{e:sconditionw1}
\langle u_0, v_1\rangle - \langle L_{01}u_0, w_0 \rangle=0.
\end{equation}
To simplify this condition, we recall the relation $L_{0}u^0(\x;
\mu)=0$. By inserting the expansions (\ref{e:uexpandcase2a}) and
(\ref{e:L0L1expandcase2}) for $L_0$ and $u^0(\x; \mu)$ into this
relation and collecting terms of $O(\mu-\mu_0)$, we get
\begin{equation} \label{f:L01u0}
L_{01}u_0=-L_{00}u_1.
\end{equation}
When this relation and the solutions of $w_0$ and $v_1$ are
substituted into the solvability condition (\ref{e:sconditionw1})
and after simple algebra, this solvability condition then reduces to
\begin{equation} \label{e:sconditionw1b}
c_0\lambda_0^2\langle u_0, L_{10}^{-1}u_0\rangle=-\langle u_0,
\widehat{v}_1\rangle-\langle u_1, \psi\rangle.
\end{equation}
From Eq. (\ref{f:P0case2}) and conditions (\ref{c:pitchfork2}),
$\langle u_0, L_{10}^{-1}u_0 \rangle\ne 0$. In addition, we will see
from Eq. (\ref{f:lambda02}) later that $\lambda_0^2\ne 0$ as well.
Thus the solvability condition (\ref{e:sconditionw1b}) yields a
unique real value for $c_0$ as
\begin{equation}
c_0=-\frac{\langle u_0, \widehat{v}_1\rangle+ \langle u_1,
\psi\rangle}{\lambda_0^2\langle u_0, L_{10}^{-1}u_0\rangle}.
\end{equation}
Consequently, the $w_0$ and $v_1$ solutions are now fully determined
and are both real. In addition, with this $c_0$ value, the $w_1$
equation (\ref{e:L00w1}) is solvable, and its solution is
\begin{equation}
w_1=\widehat{w}_1+\lambda_1L_{00}^{-1}\psi+c_1u_0,
\end{equation}
where $\widehat{w}_1$ is a real and localized particular solution to
the $w_1$ equation (\ref{e:L00w1}) but without the $\lambda_1$ term
on its right hand side, and $c_1$ is a constant. The eigenvalue
coefficient $\lambda_1$ and the constant $c_1$ will be determined
from the solvability conditions of the $v_2$ and $w_2$ equations
(\ref{e:L10v2})-(\ref{e:L00w2}).

Next we use the method of induction to show that all higher-order
terms in the perturbation series expansions
(\ref{e:vexpand})-(\ref{e:lambdaexpand}) of $(v, w, \lambda)$ can be
successively determined and are all real-valued. Suppose the $v_0,
v_1, \dots, v_n$ and $w_0, w_1, \dots, w_{n-1}$ solutions have been
fully obtained and are all real. In addition, suppose the $v_n$
solution is of the form
\begin{equation} \label{f:vn}
v_n=\widehat{v}_n+\lambda_0^2c_{n-1}L_{10}^{-1}u_0,
\end{equation}
where $\widehat{v}_n$ is a real and localized function, and
$c_{n-1}$ is a real constant. Furthermore, suppose the $w_n$
solution is of the form
\begin{equation} \label{f:wn}
w_n=\widehat{w}_n+\lambda_nL_{00}^{-1}\psi+c_nu_0,
\end{equation}
where $\widehat{w}_n$ is a known real and localized function but the
coefficients $\lambda_n$ and $c_n$ are not known yet. These
assumptions are satisfied when $n=1$ (see above), and we now show if
they hold for $n$ then they would still hold for $n+1$ as well. To
determine $\lambda_n$, we use the solvability condition of the
$v_{n+1}$ equation (\ref{e:L10vnp1}). Inserting (\ref{f:wn}) into
this solvability condition, we readily find that
\begin{equation}
\lambda_n=\frac{\langle \psi,
\sum_{k=1}^{n+1}L_{1k}v_{n+1-k}\rangle-\lambda_0^2 \langle \psi,
\widehat{w}_n+\sum_{k=1}^n \lambda_k
w_{n-k}\rangle}{\lambda_0^2\langle \psi, L_{00}^{-1}\psi\rangle},
\end{equation}
which is real. For this $\lambda_n$ value, the $v_{n+1}$ equation
(\ref{e:L10vnp1}) is solvable, and its solution is of the form
\begin{equation} \label{f:vnp1}
v_{n+1}=\widehat{v}_{n+1}+\lambda_0^2c_{n}L_{10}^{-1}u_0,
\end{equation}
where $\widehat{v}_{n+1}$ is a real and localized particular
solution to the $v_{n+1}$ equation but without the $c_n$ term in
$w_n$ on its right hand side (see (\ref{f:wn})). Notice that this
$v_{n+1}$ solution is of the same form as $v_n$ in (\ref{f:vn}) but
with the index $n$ changed to $n+1$.

To determine the constant $c_n$ in the above $w_n$ and $v_{n+1}$
solutions, we use the solvability condition of the $w_{n+1}$
equation (\ref{e:L00wnp1}), which is that its right hand side be
orthogonal to the homogeneous solution $u_0$. Inserting the above
$w_n$ and $v_{n+1}$ solutions into this solvability condition and
utilizing the relation (\ref{f:L01u0}), we find that this
solvability condition yields the $c_n$ value as
\begin{eqnarray}
\hspace{-0.5cm} c_n &  \hspace{-0.2cm} = & \hspace{-0.2cm}
\frac{1}{\lambda_0^2\langle u_0, L_{10}^{-1}u_0\rangle} \left[
\langle u_0,
L_{01}\widehat{w}_n+\sum_{k=2}^{n+1}L_{0k}w_{n+1-k}\rangle  \right.
\nonumber
\\
&& \left.  \hspace{0.5cm}  -\langle u_0,
\widehat{v}_{n+1}+\sum_{k=1}^{n}\lambda_kv_{n+1-k}\rangle-\lambda_n\langle
u_1,\psi\rangle \right],
\end{eqnarray}
which is a real constant. With this $c_n$ value, the $w_n$ and
$v_{n+1}$ solutions are now fully determined and are both real. In
addition, the $w_{n+1}$ equation (\ref{e:L00wnp1}) is now solvable,
and its general solution is
\begin{equation}
w_{n+1}=\widehat{w}_{n+1}+\lambda_{n+1}L_{00}^{-1}\psi+c_{n+1}u_0,
\end{equation}
where $\widehat{w}_{n+1}$ is a real and localized particular
solution to the $w_{n+1}$ equation (\ref{e:L00wnp1}) but without the
$\lambda_{n+1}$ term on its right hand side, and $c_{n+1}$ is a
constant. This $w_{n+1}$ solution is of the same form as $w_n$ in
(\ref{f:wn}) but with the index $n$ changed to $n+1$. This completes
the induction process.

It is noted that in the above construction of the bifurcated
eigenmodes, while $\lambda_0$ has two solutions $\lambda_{0\pm}$
(with $\lambda_{0+}=-\lambda_{0-}$) in view of Eq.
(\ref{f:lambda0}), the higher coefficients $\lambda_1, \lambda_2,
\dots$ as well as all $(v_n, w_n)$ functions in the expansions
(\ref{e:vexpand})-(\ref{e:lambdaexpand}) depend on $\lambda_0^2$
only and are thus the same for both values of $\lambda_{0\pm}$. It
is then clear that the above construction yields two eigenmodes
$(v^0_\pm, w^0_\pm, \lambda^0_\pm)$ which correspond to the two
choices of the $\lambda_0$ values, and these two eigenmodes are
related as
\begin{equation}
\lambda^0_+=-\lambda^0_-, \quad v^0_+=v^0_-, \quad w^0_+=-w^0_-.
\end{equation}
In addition, $v^0_\pm$ is always real, and $w^0_\pm$,
$\lambda^0_\pm$ are either real or purely imaginary. The asymptotic
formula for the eigenvalues $\lambda^0_\pm$ is
\begin{equation} \label{f:lambdacase2aa}
(\lambda^0)^2 \to -\lambda_0^2 \hspace{0.06cm} (\mu-\mu_0), \quad
\mu \to \mu_0,
\end{equation}
where $\lambda_0^2$ is given in (\ref{f:lambda0}).

Finally, we simplify the $\lambda_0^2$ formula (\ref{f:lambda0}) and
show that $-\lambda_0^2$ is equal to $\alpha$ as given in Eq.
(\ref{f:alpha2}). To do so, we expand operator $L_1$ in (\ref{e:L1})
around $\mu=\mu_0$. Recalling the notations (\ref{n:G}), we readily
find $L_{11}$ in the expansion (\ref{e:L0L1expandcase2}) of $L_1$ as
\begin{equation}  \label{f:L11case2}
L_{11}=G_2u_1-1,
\end{equation}
where $u_1$ is the $O(\mu-\mu_0)$ term in $u^0(\x; \mu)$'s expansion
(\ref{e:uexpandcase2a}). The expression for $u_1$ has been obtained
in Ref.~\cite{Yang_classification} as
\begin{equation}
u_1=L_{10}^{-1}u_0+b_1\psi,
\end{equation}
where $b_1$ is a real constant. Inserting the above two expressions
into (\ref{f:lambda0}) and recalling the condition of $\langle G_2,
\psi^3\rangle=0$ for pitchfork bifurcations in Theorem 1 as well as
the definition of constant $R$ in Eq. (\ref{def:R}), we get a more
explicit formula for $\lambda_0^2$ as
\begin{equation} \label{f:lambda02}
\lambda_0^2=-\frac{R} {\langle \psi, L_{00}^{-1}\psi\rangle}.
\end{equation}
In view of the conditions for pitchfork bifurcations in Theorem 1,
we see that $\lambda_0^2\ne 0$ as was mentioned before. Inserting
this $\lambda_0^2$ formula into (\ref{f:lambdacase2aa}), the final
asymptotic expression (\ref{f:lambdacase2a}) for the bifurcated
eigenvalues $\lambda^0(\mu)$ is then derived, with the constant
$\alpha$ given by Eq. (\ref{f:alpha2}).

\textbf{(b) Eigenvalue bifurcation along the bifurcated solution
branches}

Now we calculate eigenvalue bifurcations along the bifurcated
solution branches $u^\pm(\x; \mu)$. For convenience, we assume that
these bifurcated solutions appear on the right side of $\mu=\mu_0$
(when $RS>0$, see Theorem 1). The other case of these bifurcated
solutions appearing on the left side of $\mu=\mu_0$ can be similarly
treated with trivial modifications, and both cases yield the same
eigenvalue formula given in Eq. (\ref{f:lambdapmcase2b}).

The bifurcated solitary waves near the bifurcation point $\mu=\mu_0$
have the following perturbation series
expansion~\cite{Yang_classification}
\begin{equation} \label{e:uexpandcase2b}
u^\pm(\x; \mu) = \sum_{k=0}^\infty (\mu-\mu_0)^{k/2}u_k(\x).
\end{equation}
Since these solutions are assumed to exist on the right side of
$\mu=\mu_0$, all functions $u_1, u_2, \dots$ in this expansion are
real-valued. Operators $L_0$ and $L_1$ on these bifurcated solution
branches are expanded as
\begin{equation}  \label{e:L0L1expandcase2b}
L_0^\pm=\sum_{k=0}^\infty (\mu-\mu_0)^{k/2} L_{0k}, \quad L_1^\pm
=\sum_{k=0}^\infty (\mu-\mu_0)^{k/2} L_{1k},
\end{equation}
and all terms in these expansions are real-valued too. Note that
quantities $u_k, L_{0k}, L_{1k}, k=1, 2, \dots$ in these expansions
are different from those in previous expansions
(\ref{e:uexpandcase2a})-(\ref{e:L0L1expandcase2}). The
linear-stability eigenmodes $(v, w, \lambda)$ bifurcated from the
zero eigenmode (\ref{e:zeromode2}) now have the perturbation series
expansions
\begin{eqnarray}
&& \hspace{-0.4cm} v^\pm(\x;\mu)=\sum_{k=0}^\infty (\mu-\mu_0)^{k/2} v_k(\x),   \label{e:vexpandb} \\
&& \hspace{-0.4cm} w^\pm(\x; \mu)  = \lambda_0 \sum_{k=0}^\infty
(\mu-\mu_0)^{k/2}
w_k(\x),   \label{e:wexpandb} \\
&&  \hspace{-0.4cm}
\lambda^\pm(\mu) =
i\lambda_0(\mu-\mu_0)^{1/2}\left(1+\sum_{k=1}^\infty
(\mu-\mu_0)^{k/2} \lambda_k\right). \label{e:lambdaexpandb}
\end{eqnarray}
Below we construct these perturbation series solutions to all
orders.

Before this construction, we first derive a few relations on
functions $u_0, u_1, u_2$ and $u_3$ in (\ref{e:uexpandcase2b}),
which will be needed in later calculations. By inserting expansions
(\ref{e:uexpandcase2b}) and (\ref{e:L0L1expandcase2b}) into
equations (\ref{e:L0u}) and (\ref{e:L1umu}) and at suitable orders,
we get the following relations
\begin{eqnarray}
&& L_{01}u_0=-L_{00}u_1,    \label{f:L01u0b}  \\
&& L_{02}u_0=-L_{01}u_1-L_{00}u_2,   \label{f:L02u0b}  \\
&& L_{11}u_1=2(u_0-L_{10}u_2),   \label{f:L11u1b}  \\
&& L_{12}u_1=2u_1-2L_{11}u_2-3L_{10}u_3.   \label{f:L12u1b}
\end{eqnarray}
In addition,
\begin{equation} \label{e:u1case2b}
u_1=b_1\psi,
\end{equation}
where $b_1=\pm \sqrt{6R/S}$ which is non-zero
\cite{Yang_classification}. Notice that $b_1$ is also real-valued
here since $RS>0$ by our earlier assumption.

We now substitute the perturbation expansions
(\ref{e:vexpandb})-(\ref{e:lambdaexpandb}) into the linear-stability
eigenvalue problem (\ref{e:L0wL1v}). From various orders of
$(\mu-\mu_0)^{1/2}$, we get a sequence of linear equations for
$(v_k, w_k)$ as
\begin{eqnarray}
&& \hspace{-0.5cm}  L_{10}v_0=0,    \label{e:L10v0b} \\
&& \hspace{-0.5cm}  L_{00}w_0=0,   \label{e:L00w0b}  \\
&& \hspace{-0.5cm}  L_{10}v_1=\lambda_0^2w_0-L_{11}v_0,   \label{e:L10v1b}  \\
&& \hspace{-0.5cm}  L_{00}w_1=v_0-L_{01}w_0, \label{e:L00w1b} \\
&& \hspace{-0.5cm}  L_{10}v_2=\lambda_0^2(w_1+\lambda_1w_0)-(L_{11}v_1+L_{12}v_0),   \label{e:L10v2b}  \\
&& \hspace{-0.5cm}
L_{00}w_2=v_1+\lambda_1v_0-(L_{01}w_1+L_{02}w_0),\label{e:L00w2b}
\\
&& \hspace{-0.5cm} \dots\dots\dots  \nonumber \\
&& \hspace{-0.5cm}
L_{10}v_{n+1}=\lambda_0^2\left(w_n+\sum_{k=1}^n \lambda_k w_{n-k}\right)-\sum_{k=1}^{n+1}L_{1k}v_{n+1-k},  \label{e:L10vnp1b} \\
&& \hspace{-0.5cm}
L_{00}w_{n+1}=v_{n}+\sum_{k=1}^{n}\lambda_kv_{n-k}-\sum_{k=1}^{n+1}L_{0k}w_{n+1-k},
\label{e:L00wnp1b}
\\
&& \hspace{-0.5cm} \dots\dots\dots  \nonumber
\end{eqnarray}
In view of the assumption in Theorem 3, the solution to the $v_0$
equation (\ref{e:L10v0b}), after eigenfunction rescaling, can be
taken as
\begin{equation}
v_0=u_1,
\end{equation}
where $u_1$ is given in (\ref{e:u1case2b}). The solution to the
$w_0$ equation (\ref{e:L00w0b}) is
\begin{equation}
w_0=c_0u_0,
\end{equation}
where $c_0$ is a constant to be determined. When these $(v_0, w_0)$
solutions are inserted into the $(v_1, w_1)$ equations
(\ref{e:L10v1b})-(\ref{e:L00w1b}) and relations (\ref{f:L01u0b}),
(\ref{f:L11u1b}) utilized, we find that the solvability conditions
of the $(v_1, w_1)$ equations are automatically satisfied due to the
condition of $\langle u_0, \psi\rangle=0$ for pitchfork bifurcations
in Theorem 1, and these $(v_1, w_1)$ equations admit localized
solutions of the form
\begin{equation}
v_1=2(u_2-L_{10}^{-1}u_0)+c_0\lambda_0^2 L_{10}^{-1}u_0,
\end{equation}
and
\begin{equation}
w_1=L_{00}^{-1}u_1+c_0u_1+c_1u_0,
\end{equation}
where $L_{10}^{-1}u_0$ and $L_{00}^{-1}u_1$ are localized functions
and $c_1$ is another constant. It is noted that the homogeneous
solution (in proportion to $u_1$) to the $v_1$ equation is not
included in the above $v_1$ solution since this term can be lumped
into the $v_0$ term and then eliminated by a rescaling of the
eigenfunction $(v, w)$ --- the same treatment we have applied
previously in case (a) (see Eq. (\ref{f:v1})).

Now we consider the $(v_2, w_2)$ equations
(\ref{e:L10v2b})-(\ref{e:L00w2b}). Inserting the above $(v_0, v_1,
w_0, w_1)$ solutions into the right hand side of the $v_2$ equation,
utilizing the relations (\ref{f:L01u0b})-(\ref{f:L12u1b}) and after
simple algebra, the solvability condition of this $v_2$ equation,
which requires that its right hand side be orthogonal to the
homogeneous solution $u_1$, yields
\begin{equation} \label{e:lambda02a}
\lambda_0^2 = (2-c_0\lambda_0^2) \frac{\langle u_1,
u_1\rangle+2\langle u_0, u_2\rangle-2\langle u_0, L_{10}^{-1}u_0
\rangle}{\langle u_1, L_{00}^{-1}u_1\rangle}.
\end{equation}
It is noted that $u_1$ is proportional to $\psi$ and is nonzero, see
Eq. (\ref{e:u1case2b}). Thus due to the conditions
(\ref{c:pitchfork2}), the denominator in the above equation is
nonzero, i.e., $\langle u_1, L_{00}^{-1}u_1\rangle\ne 0$. From the
expansion (\ref{e:uexpandcase2b}) of the solutions $u^\pm(\x; \mu)$
and the condition $\langle u_0, \psi\rangle=0$, we see that the
expansion for the power function $P_\pm(\mu)$ is
\begin{equation}
P_\pm(\mu)=\langle u_0, u_0\rangle + P_\pm'(\mu_0)
(\mu-\mu_0)+O[(\mu-\mu_0)^{3/2}],
\end{equation}
where
\begin{equation}
P_\pm'(\mu_0) =\langle u_1, u_1\rangle+2\langle u_0, u_2\rangle
\end{equation}
is the power slope at the bifurcation point $\mu=\mu_0$. From Eq.
(\ref{f:P1case2}) in Theorem 2, we also know that
\begin{equation}
P_+'(\mu_0)=P_-'(\mu_0)=2\langle u_0,
L_{10}^{-1}u_0\rangle+\frac{6R^2}{S}.
\end{equation}
Using these relations as well as Eq. (\ref{e:u1case2b}), the
solvability condition (\ref{e:lambda02a}) of the $v_2$ equation
reduces to
\begin{equation}  \label{e:lambda02temp}
\lambda_0^2 = (2-c_0\lambda_0^2) \,  \frac{R}{\langle \psi,
L_{00}^{-1}\psi\rangle}.
\end{equation}
From the conditions for pitchfork bifurcations in Theorem 1, $R\ne
0$, thus $\lambda_0^2\ne 0$.

Carrying similar calculations to the $w_2$ equation (\ref{e:L00w2b})
and recalling the formula (\ref{f:P0case2}), the solvability
condition of this $w_2$ equation yields
\begin{equation}  \label{r:w2solb}
2-c_0\lambda_0^2
= \frac{2P_\pm'(\mu_0)}{P_0'(\mu_0)}.
\end{equation}
When this equation is inserted into (\ref{e:lambda02temp}), an
expression for $\lambda_0^2$ is then obtained as
\begin{equation} \label{f:lambda0case2b}
\lambda_0^2=\frac{2P_\pm'(\mu_0)}{P_0'(\mu_0)} \,  \frac{R}{\langle
\psi, L_{00}^{-1}\psi\rangle},
\end{equation}
which is real and nonzero. Inserting this $\lambda_0^2$ formula into
(\ref{r:w2solb}), the constant $c_0$ can be obtained and is also
real.

When $\lambda_0$ and $c_0$ are given by Eqs.
(\ref{r:w2solb})-(\ref{f:lambda0case2b}), the solvability conditions
of the $(v_2, w_2)$ equations (\ref{e:L10v2b})-(\ref{e:L00w2b}) are
satisfied. Utilizing the relation (\ref{f:L01u0b}), the $(v_2, w_2)$
solutions are of the form
\begin{equation}
v_2=\widehat{v}_2+\lambda_0^2(c_1+\lambda_1c_0)L_{10}^{-1}u_0,
\end{equation}
and
\begin{equation}
w_2=\widehat{w}_2+\lambda_1 L_{00}^{-1}u_1+c_1u_1+c_2u_0,
\end{equation}
where $\widehat{v}_2$ and $\widehat{w}_2$ are localized functions
which satisfy equations (\ref{e:L10v2b})-(\ref{e:L00w2b}) but
without the $c_1$ and $\lambda_1$ terms on their right hand sides,
and $c_2$ is another constant. These constants $\lambda_1, c_1$ and
$c_2$ will be determined from the solvability conditions of the
higher $(v_n, w_n)$ equations.

Using the method of induction and after straightforward algebra, we
can show that the perturbation series solution
(\ref{e:vexpandb})-(\ref{e:lambdaexpandb}) for the eigenmode
$(v,w,\lambda)$ can be determined to all orders. In addition, the
$(v_n, w_n)$ terms for $n\ge 2$ are of the form
\begin{equation}
v_n=\widehat{v}_n+\lambda_0^2(c_{n-1}+\lambda_{n-1}c_0)L_{10}^{-1}u_0,
\quad n\ge 2,
\end{equation}
and
\begin{equation}
w_n=\widehat{w}_n+\lambda_{n-1} L_{00}^{-1}u_1+c_{n-1}u_1+c_nu_0,
\quad n\ge 2,
\end{equation}
where $\widehat{v}_n$ and $\widehat{w}_n$ are certain localized
functions, and $\lambda_{n-1}, c_{n-1}, c_n$ are constants. Since
this induction calculation is analogous to that for case (a) in the
earlier text, the details are omitted here.

In the above construction of the bifurcated eigenmodes, since
$\lambda_0$ has two solutions from Eq. (\ref{f:lambda0case2b}), two
eigenmodes are then obtained. The eigenvalues of these two modes are
either real or purely imaginary and are opposite of each other. From
the perturbation expansion of these eigenvalues in Eq.
(\ref{e:lambdaexpandb}) as well as Eq. (\ref{f:lambda0case2b}), we
see that the asymptotic formula for these eigenvalues near the
bifurcation point is
\begin{equation}
(\lambda^\pm)^2 \to -\frac{2P_\pm'(\mu_0)}{P_0'(\mu_0)} \,
\frac{R}{\langle \psi, L_{00}^{-1}\psi\rangle} \,  (\mu-\mu_0),
\quad \mu\to \mu_0,
\end{equation}
which is the same as the formula (\ref{f:lambdapmcase2b}) in Theorem
3. This completes the proof of Theorem 3. $\Box$

\vspace{0.1cm} \textbf{Proof of Theorem 4} \ From the assumptions in
Theorem 4, the solitary wave $u_0(\x)$ at the bifurcation point
$\mu=\mu_0$ is linearly stable; and when $0<|\mu-\mu_0|\ll 1$, the
only instability-inducing eigenvalue bifurcation is from the origin.
In Theorem 3, we have shown that from the origin, a single pair of
eigenvalues $\pm \lambda$ in ${\cal L}$ bifurcate out along the real
or imaginary axis. Thus the linear stability of these solitary waves
near $\mu=\mu_0$ is determined entirely by whether this pair of
eigenvalues are real or purely imaginary. On the smooth solitary
wave branch $u^0(\x; \mu)$, this pair of eigenvalues are given
asymptotically by Eq. (\ref{f:lambdacase2a}). Thus if $\alpha>0$,
these eigenvalues are real when $\mu>\mu_0$ and imaginary when
$\mu<\mu_0$, hence the solitary waves are linearly unstable when
$\mu>\mu_0$ and linearly stable when $\mu<\mu_0$. If $\alpha<0$, the
situation is just the opposite. In both cases, stability switches at
the bifurcation point. On the bifurcated solitary branches
$u^\pm(\x; \mu)$, the bifurcated eigenvalues are given
asymptotically by Eq. (\ref{f:lambdapmcase2b}). The formula
(\ref{f:beta}) for the constant $\beta$ shows that when the two
power slopes $P_0'(\mu_0)$ and $P_\pm'(\mu_0)$ have the same sign,
$\beta$ and $\alpha$ would have the opposite sign, meaning that
eigenvalues for the $u^\pm(\x; \mu)$ and $u^0(\x; \mu)$ branches
bifurcate along perpendicular directions from the origin, hence
these solution branches have opposite linear stability. On the other
hand, if the two power slopes $P_0'(\mu_0)$ and $P_\pm'(\mu_0)$ have
the opposite sign, $\beta$ and $\alpha$ would have the same sign,
hence the $u^\pm(\x; \mu)$ and $u^0(\x; \mu)$ branches would have
the same linear stability. This completes the proof of Theorem 4.
$\Box$

\textbf{Proof of Theorem 5} \ If zero is the $n$-th largest discrete
eigenvalue of $L_{10}$ with $n\ge 3$, then $L_{10}$ has two positive
eigenvalues, hence the positive solitary wave $u_0(\x)$ at the
bifurcation point $\mu=\mu_0$ is linearly unstable by the
generalized Vakhitov-Kolokolov stability criterion \cite{Yang_SIAM}.
If zero is the second-largest discrete eigenvalue of $L_{10}$, then
$L_{10}$ has one positive eigenvalue. Meanwhile, by the conditions
of pitchfork bifurcations in Theorem 1, $\langle u_0, \psi\rangle
=0$. In this case, if $P_0'(\mu_0)<0 \ (>0)$, then the positive
solitary wave $u_0(\x)$ is linearly unstable (stable) by the
generalized Vakhitov-Kolokolov stability criterion \cite{Yang_SIAM}.
When $u_0(\x)$ is linearly unstable, solitary waves near the
bifurcation point $\mu=\mu_0$ are clearly also linearly unstable.

Next we consider the case when $P_0'(\mu_0)>0$, $P_\pm'(\mu_0)\ne 0$
and zero is the second largest and simple discrete eigenvalue of
$L_{10}$. In this case, $u_0(\x)$ at the bifurcation point is
linearly stable (see above). In addition, for positive solitary
waves, all eigenvalues in ${\cal L}$ are real or purely imaginary
(see Ref. \cite{Yang_SIAM}, Theorem 5.2), thus no complex
eigenvalues can bifurcate out when $\mu\ne \mu_0$. So the
assumptions in Theorem 4 are satisfied. Since $u_0(\x)$ is positive,
zero is then the largest eigenvalue of $L_{00}$ and is simple
\cite{Struwe}, and $\langle \psi, L_{00}^{-1}\psi\rangle<0$ as
$L_{00}$ is semi-negative definite and $\psi\ne 0$. In addition,
zero is a simple eigenvalue of $L_{10}$, $P_0'(\mu_0)\ne 0$ and
$P_\pm'(\mu_0)\ne 0$ by our assumptions above. Thus the conditions
of Theorem 4 are also met. Hence Theorem 4 can be applied. Using
this theorem, together with Remark 3 and the fact of $\langle \psi,
L_{00}^{-1}\psi\rangle<0$, we can then prove the results in the four
cases of Theorem 5 as below.
\begin{enumerate}
\item When the bifurcated solitary waves $u^\pm(\x; \mu)$ appear on the right side of
$\mu=\mu_0$ and $P_\pm'(\mu_0) < P_0'(\mu_0)$, $\alpha$ is positive
by Remark 3. Then by Theorem 4, the smooth solution branch $u^0(\x;
\mu)$ is stable for $\mu<\mu_0$ and unstable for $\mu>\mu_0$.
Regarding the bifurcated branches, they and the unstable smooth
branch (on the right side of $\mu=\mu_0$) should have the opposite
(same) stability when $P_0'(\mu_0)$ and $P_\pm'(\mu_0)$ have the
same (opposite) sign. Since $P_0'(\mu_0)>0$, these bifurcated
branches are then stable when $P_\pm'(\mu_0)>0$ and unstable when
$P_\pm'(\mu_0) <0$.

\item When the bifurcated solitary waves appear on the right side of
$\mu=\mu_0$ and $P_\pm'(\mu_0) > P_0'(\mu_0)>0$, $\alpha$ is
negative by Remark 3. Thus the smooth solution branch is unstable
for $\mu<\mu_0$ and stable for $\mu>\mu_0$. Since $P_\pm'(\mu_0)$
and $P_0'(\mu_0)$ are now both positive, the bifurcated branches
then have the opposite stability of the stable smooth branch (on the
right side of $\mu=\mu_0$) are thus always unstable.

\item When the bifurcated solitary waves appear on the left side of
$\mu=\mu_0$ and $P_\pm'(\mu_0) < P_0'(\mu_0)$, $\alpha$ is negative
by Remark 3. Then by Theorem 4, the smooth solution branch is
unstable for $\mu<\mu_0$ and stable for $\mu>\mu_0$. The bifurcated
branches (with $\mu<\mu_0$) are stable (opposite of the unstable
smooth branch) if $P_\pm'(\mu_0) >0$ and unstable (same as the
unstable smooth branch) if $P_\pm'(\mu_0) <0$.

\item When the bifurcated solitary waves appear on the left side of
$\mu=\mu_0$ and $P_\pm'(\mu_0) > P_0'(\mu_0)>0$, $\alpha$ is
positive by Remark 3; hence by Theorem 4, the smooth solution branch
is stable for $\mu<\mu_0$ and unstable for $\mu>\mu_0$. Since both
$P_\pm'(\mu_0)$ and $P_0'(\mu_0)$ are now positive, the bifurcated
branches are always unstable (opposite of the stable smooth branch
on the left side of $\mu=\mu_0$). This completes the proof of
Theorem 5. $\Box$
\end{enumerate}

\section{Numerical examples}
In this section, we use a few numerical examples to illustrate and
confirm the above analytical stability results. These examples
contain both positive and sign-indefinite solitary waves under
various self-focusing and self-defocusing nonlinearities in one and
higher spatial dimensions.

\textbf{Example 1} \ Our first example is the one-dimensional GNLS
equation (\ref{e:U}) with a symmetric double-well potential and
cubic-quintic nonlinearity,
\begin{equation}  \label{e:Uexample}
iU_t+U_{xx}-V(x)U+|U|^2U-0.25 |U|^4U=0,
\end{equation}
where the symmetric double-well potential $V(x)$ is taken as
\begin{equation}  \label{e:potential_example1}
V(x)=-2.8\left[\mbox{sech}^2(x+1.5)+\mbox{sech}^2(x-1.5)\right]
\end{equation}
and is shown in Fig. 3(a). Pitchfork bifurcations of positive
solitary waves in this equation have been reported in
\cite{Yang_classification}, and the power diagram for these
bifurcations is displayed in Fig. 3(b). This power diagram shows
that two pitchfork bifurcations occur at points `A,B' of the power
diagram. At positions `c,d,e', profiles of solitary waves are shown
in Fig. 3(c,d,e) respectively. It is seen that solitary waves at
positions `c,d' are symmetric, while solitary waves at position `e'
are asymmetric. Thus symmetry-breaking bifurcations occur at both
`A,B' points. In particular, the c-d power branch is the symmetric
(smooth) branch, and the e-branch is the asymmetric (bifurcated)
branch.

Since solitary waves in this example are positive, Theorem 5
applies. It is known that for both pitchfork bifurcations at points
`A,B', zero is the second largest eigenvalue of $L_{10}$
\cite{Yang_classification}. In addition, the power diagram in Fig.
3(b) shows that at both `A,B' points, $0<P_\pm'(\mu_0)<P_0'(\mu_0)$.
Thus Theorem 5 predicts that near the bifurcation point `A', the
smooth (symmetric) branch is linearly stable on the left side of `A'
and unstable on the right side of `A', and the bifurcated
(asymmetric) branch (which is on the right side of `A') is linearly
stable. Regarding the bifurcation point `B', Theorem 5 predicts that
the symmetric branch is unstable on the left side of `B' and stable
on the right side of `B', and the bifurcated asymmetric branch
(which is on the left side of `B') is linearly stable.

\begin{figure}[h!]
\centerline{\includegraphics[width=0.65\textwidth]{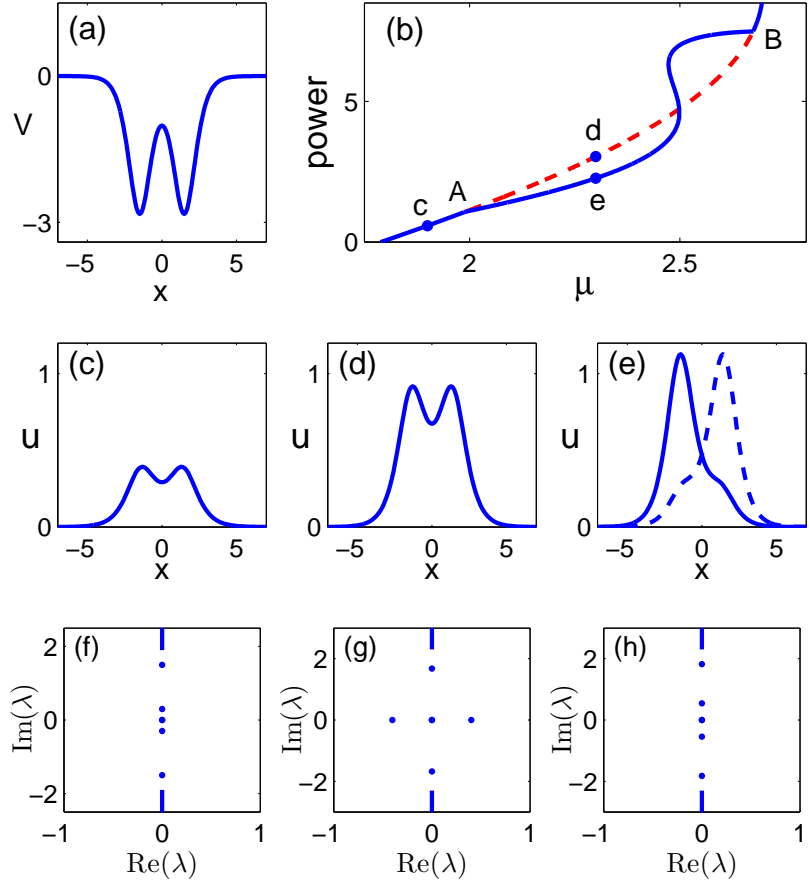}}
\caption{Pitchfork bifurcations of solitary waves and their
linear-stability behaviors in Example 1. (a) The potential
(\ref{e:potential_example1}); (b) the power diagram (solid blue and
dashed red indicate stable and unstable solutions respectively);
(c,d,e) profiles of solitary waves at positions of the same letters
in (b); the two asymmetric solitary waves in (e) are mirror images
of each other with respect to $x$; (f,g,h) linear-stability spectra
of the solitary waves in (c,d,e) respectively. } \label{f:fig3}
\end{figure}

Numerically we have found that these analytical predictions are
entirely correct. Specifically, we have found numerically that the
segment of the symmetric branch between points `A,B' is linearly
unstable, and the other segments/branches of the power diagram are
all linearly stable. This stability information is indicated by the
solid blue and dashed red lines in Fig. 3(b) for stable and unstable
parts respectively. To support this stability result, the
linear-stability spectra for solitary waves at locations `c,d,e' of
the power diagram are numerically computed by the Fourier
collocation method \cite{Yang_SIAM}, and the results are displayed
in Fig. 3(f,g,h). It is seen that the spectrum at `d' contains a
positive eigenvalue, hence its symmetric solitary wave is linearly
unstable. The spectra at `c,e', on the other hand, lie entirely on
the imaginary axis, thus those solitary waves are linearly stable.
These numerical stability results agree completely with the above
analytical predictions.

The reader may notice that the asymmetric `e'-branch in Fig. 3(b)
contains two additional saddle-node (fold) bifurcations. As was
explained in \cite{Yang_saddle1,Yang_saddle2}, there is no stability
switching at saddle-node bifurcations in the GNLS equations
(\ref{e:U}), thus it is not surprising that the entire asymmetric
`e'-branch in Fig. 3(b) is linearly stable despite these saddle-node
bifurcations.

\textbf{Example 2} \ Our second example is the one-dimensional GNLS
equation (\ref{e:U}) with self-focusing cubic nonlinearity and a
periodic potential,
\begin{equation}
iU_t+U_{xx}-V(x)U+|U|^2U=0,
\end{equation}
where the potential $V(x)$ is
\begin{equation}
V(x)=6\sin^2 \hspace{-0.05cm} x.
\end{equation}
This equation admits a family of sign-indefinite solitary waves of
the form (\ref{e:Usoliton}) in the semi-infinite bandgap
\cite{Akylas_2012}, and its power diagram is shown in Fig. 4(a).
This power diagram contains three branches which are connected with
each other. At the intersection point `A' between the lower and
middle branches, a pitchfork bifurcation occurs. This pitchfork
bifurcation is better seen in Fig. 4(b), which shows an
amplification of the power diagram in Fig. 4(a) around this
intersection point. Solitary waves on the lower power branch are
anti-symmetric (see Fig. 4(c)), whereas solitary waves on the middle
power branch are asymmetric (see Fig. 4(d)), hence a
symmetry-breaking bifurcation occurs at point `A'.

At this bifurcation point `A', we have checked numerically that
$\langle \psi, L_{00}^{-1}\psi\rangle<0$. In addition, the power
diagram in Fig. 4(b) shows that the bifurcated (asymmetric) solitary
waves appear on the right side `A', and
$P_\pm'(\mu_0)>P_0'(\mu_0)>0$. Thus, Remark 3 gives $\alpha<0$. We
have also checked that the solitary wave at point `A' is linearly
stable, and near this point no complex eigenvalues appear. Then
Theorem 4 predicts that the lower branch is unstable on the left
side of `A' and stable on the right side of `A'. In addition, the
bifurcated asymmetric branch is unstable.

\begin{figure}[h!]
\centerline{\includegraphics[width=0.8\textwidth]{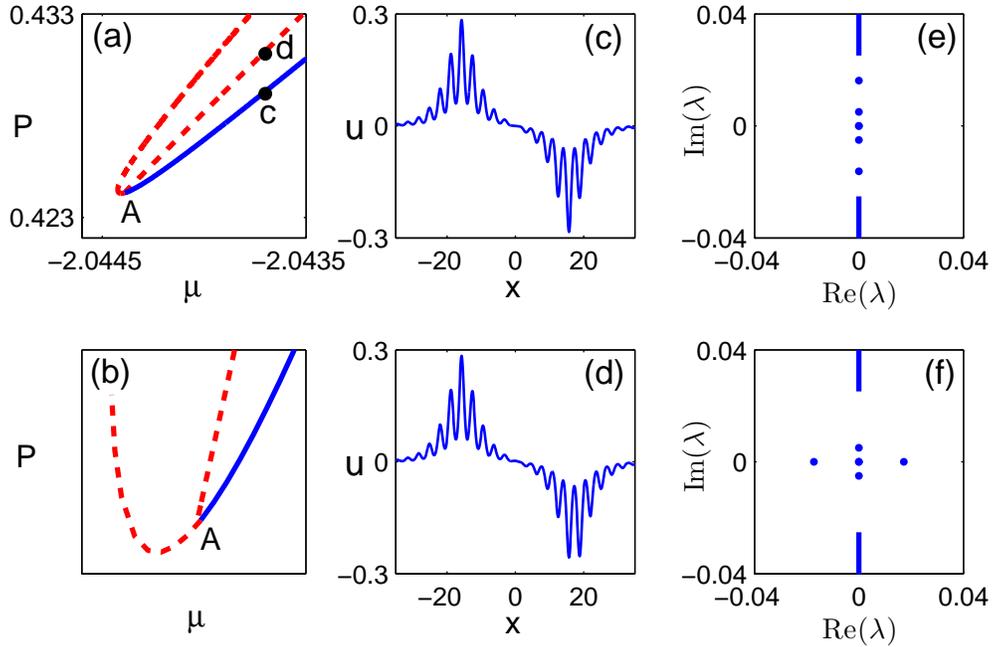}}
\caption{Pitchfork bifurcations of solitary waves and their
linear-stability behaviors in Example 2. (a) The power diagram
(solid blue for stable branches and dashed red for unstable ones);
(b) amplification of the power diagram in (a) near the pitchfork
bifurcation point `A'; (c,d) profiles of solitary waves at locations
`c,d' of the power diagram in (a); (e,f) linear-stability spectra
for solitary waves in (c,d) respectively. }  \label{f:fig4}
\end{figure}

Numerically we have found that these analytical predictions are
again all correct. Specifically, we have determined the stability of
these solitary waves through computation of their linear-stability
spectra, and the stability results are indicated on the power
diagram in Fig. 4(a,b), where the stable and unstable branches are
marked as solid blue and dashed red lines respectively. We see that
these stability results are in full agreement with the above
analytical predictions. To corroborate these stability results, the
full linear-stability spectra for solitary waves at locations `c,d'
of Fig. 4(a) are displayed in Fig. 4(e,f). These spectra confirm
that the anti-symmetric solitary wave [Fig. 4(c)] at location `c' of
the lower power branch is indeed linearly stable, whereas the
asymmetric solitary wave [Fig. 4(d)] at location `d' of the middle
branch is indeed linearly unstable. The unstable eigenvalue at
location `d' is real positive, and it bifurcates out from the
pitchfork bifurcation point `A', in agreement with our theory.

\textbf{Example 3} \ Our third example is the seventh-power GNLS
equation with a symmetric double-well potential,
\begin{equation}
iU_t+U_{xx}-V(x)U+|U|^6U=0,
\end{equation}
where the potential $V(x)$ is
\begin{equation}  \label{e:potential4}
V(x)=-3\: \mbox{sech}^2(x+1)-3\: \mbox{sech}^2(x-1),
\end{equation}
which is shown in Fig. 5(b). This equation admits a family of
positive solitary waves whose power diagram is displayed in Fig.
5(a). This power diagram shows that a pitchfork bifurcation occurs
at the point `A'. On the upper b-$c_1$ branch, solitary waves are
symmetric (see Fig. 5(b,c)), whereas on the lower $c_2$ branch,
solitary waves are asymmetric (see Fig. 5(c)).

At the pitchfork bifurcation point `A', we have checked that zero is
the second largest eigenvalue of $L_{10}$. Thus from the power
diagram in Fig. 5(a), Theorem 5 predicts that the symmetric b-$c_1$
branch is stable on the left side of `A' and unstable on the right
side of `A'. In addition, the asymmetric $c_2$ branch is unstable.
These analytical predictions fully agree with the numerical
stability results in Fig. 5(a) (where the stable and unstable
solutions are indicated). These stability results are further
corroborated in Fig. 5(d,e,f), where linear-stability spectra for
solitary waves at locations 'b, $c_1$, $c_2$' are displayed.

\begin{figure}[h!]
\centerline{\includegraphics[width=0.8\textwidth]{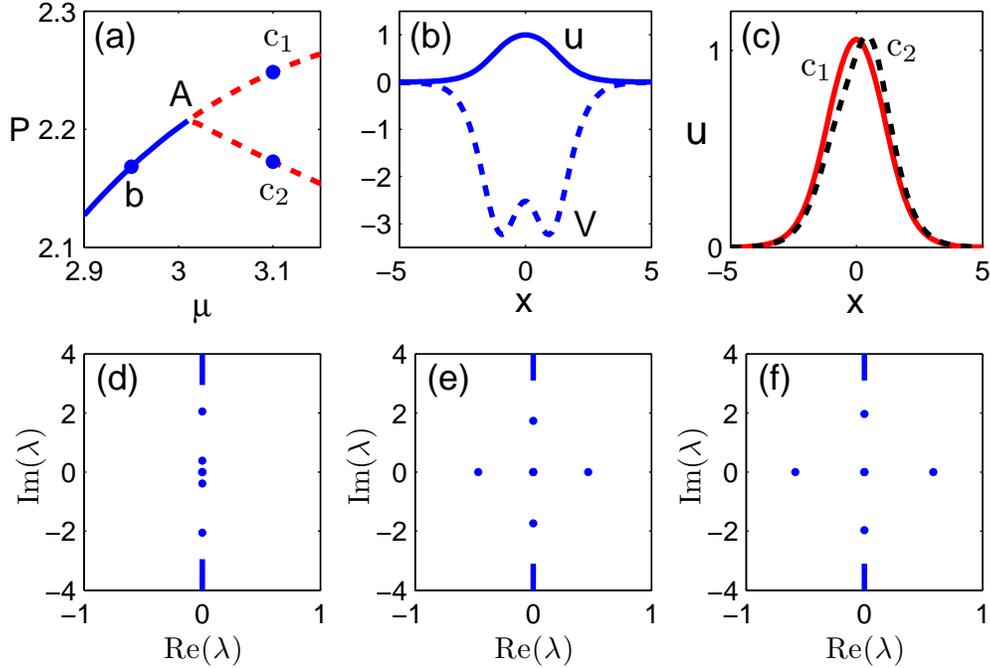}}
\caption{Pitchfork bifurcations of solitary waves and their
linear-stability behaviors in Example 3. (a) The power diagram
(solid blue is stable and dashed red unstable); (b) profiles of the
potential $V(x)$ (dashed) and the solitary wave (solid) at location
`b' of the power diagram; (c) solitary wave profiles at locations
`$c_1$' (solid) and `$c_2$' (dashed) of the power diagram; (e,f,g)
linear-stability spectra for solitary waves at locations `b, $c_1$,
$c_2$' respectively.  } \label{f:fig5}
\end{figure}

In this example, after the pitchfork bifurcation occurs (i.e., on
the right side of `A'), both the symmetric and asymmetric solution
branches are linearly unstable. A similar bifurcation was reported
numerically in \cite{Kirr_2011} for the eleventh-power nonlinearity
but was not found for the present seventh-power nonlinearity.

\textbf{Example 4} \  Our last example is the two-dimensional GNLS
equation with self-defocusing cubic nonlinearity and a symmetric
double-well potential,
\begin{equation}
iU_t+U_{xx}+U_{yy}-V(x,y)U-|U|^2U=0,
\end{equation}
where the potential $V(x,y)$ is
\begin{equation}  \label{e:2Dpotential}
V(x,y)=-6\left(e^{-[(x+1.5)^2+y^2]}+e^{-[(x-1.5)^2+y^2]}\right),
\end{equation}
which is shown in Fig. 6(a). This equation admits a family of
sign-indefinite solitary waves (\ref{e:Usoliton}) whose power
diagram is given in Fig. 6(b). It is seen that a pitchfork
bifurcation occurs at the point `A'. At positions 'd,e,f' of the
power diagram, profiles of the solitary waves are displayed in Fig.
6(d,e,f). Solitary waves at `d,e' are anti-symmetric in $x$ and
symmetric in $y$, whereas the solitary wave at `f' is asymmetric in
$x$ and symmetric in $y$. Thus this pitchfork bifurcation is also a
symmetry-breaking bifurcation.

At this bifurcation point, we have checked numerically that zero is
the largest eigenvalue of $L_{10}$, and $\langle \psi,
L_{00}^{-1}\psi\rangle>0$. In addition, the solitary wave at point
`A' is linearly stable, and the solitary waves nearby do not possess
complex eigenvalues. Thus from Remark 3 and the power diagram in
Fig. 6(b), Theorem 4 predicts that the anti-symmetric solution
branch is unstable on the left side of `A' and stable on the right
side of `A', and the bifurcated asymmetric branch is stable. These
predictions agree with our numerical stability results shown in Fig.
6(b). The numerical-stability results are further illustrated in
Fig. 6(g,h,i), where the stability spectra for solitary waves in
Fig. 6(d,e,f) are plotted. These spectra corroborate the
numerical-stability results in the power diagram of Fig. 6(b) and
support our analytical predictions.

\begin{figure}[h!]
\centerline{\includegraphics[width=0.7\textwidth]{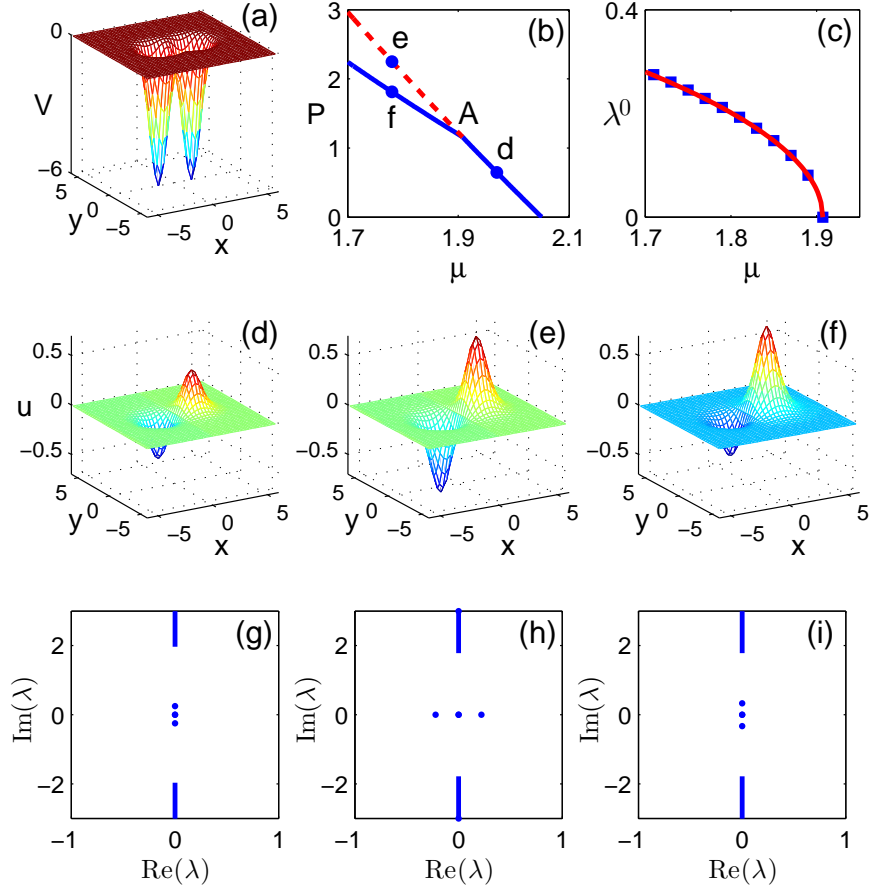}}
\caption{Pitchfork bifurcations of solitary waves and their
linear-stability behaviors in Example 4. (a) The two-dimensional
double-well potential (\ref{e:2Dpotential}); (b) the power diagram
(solid blue is stable and dashed red unstable); (c) the unstable
eigenvalue $\lambda^0$ versus $\mu$ on the anti-symmetric solution
branch of (b) [squares: numerical values; solid red line: analytical
prediction from formula (\ref{f:lambdacase2a})]; (d,e,f) profiles of
solitary waves at locations `d,e,f' of the power diagram in (b);
(g,h,i) stability spectra for solitary waves in (d,e,f)
respectively. } \label{f:fig6}
\end{figure}

In the previous examples, our comparison between analytical and
numerical stability results was qualitative. Here for this example
4, we will also make a quantitative comparison on unstable
eigenvalues in order to completely verify our eigenvalue formulae in
Theorem 3. Specifically, we notice that the anti-symmetric solution
branch in Fig. 6(b) is unstable on the left side of `A', and this
instability is induced by a positive eigenvalue which is predicated
analytically by the formula (\ref{f:lambdacase2a}) in Theorem 3.
Numerically we have determined this unstable eigenvalue $\lambda^0$
for various values of $\mu$ by the highly-accurate
Newton-conjugate-gradient method \cite{Yang_SIAM}, and these
numerical eigenvalues are plotted in Fig. 6(c) as blue squares.
Further examination of this numerical data shows that as $\mu\to
\mu_0$, where $\mu_0\approx 1.9072149$ is the propagation-constant
value at the bifurcation point `A', the numerical eigenvalue
$\lambda^0$ behaves as
\begin{equation}  \label{e:lambdanum}
(\lambda^0)^{\hspace{0.03cm} 2}_{num} \to
\alpha_{num}\hspace{0.05cm} (\mu-\mu_0), \qquad \mu\to \mu_0,
\end{equation}
where the numerical coefficient is
\[
\alpha_{num}\approx -0.3788744.
\]
In the analytical eigenvalue formula (\ref{f:lambdacase2a}), the
coefficient $\alpha$ from formula (\ref{f:alpha2}) is found to be
\[
\alpha_{anal} \approx -0.3788744.
\]
We see that the numerical eigenvalue formula (\ref{e:lambdanum}) and
the analytical formula (\ref{f:lambdacase2a}) are in complete
quantitative agreement.

On the bifurcated asymmetric solution branch in Fig. 6(b), the
solitary waves possess a pair of purely imaginary discrete
eigenvalues which are predicted analytically by the formula
(\ref{f:lambdapmcase2b}) in Theorem 3. We have quantitatively
compared the numerical values of those imaginary eigenvalues against
the analytical formula (\ref{f:lambdapmcase2b}) and found complete
agreement as well. Thus both eigenvalue formulae
(\ref{f:lambdacase2a}) and (\ref{f:lambdapmcase2b}) in Theorem 3 are
numerically verified.

\section{Summary}
In this article, linear stability of both sign-definite (positive)
and sign-indefinite solitary waves near pitchfork bifurcations has
been analyzed for the generalized nonlinear Schr\"odinger equations
(\ref{e:U}) with arbitrary forms of nonlinearity and external
potentials in arbitrary spatial dimensions. Bifurcations of
linear-stability eigenvalues associated with these pitchfork
bifurcations have been analytically calculated, and their
expressions are given by the formulae (\ref{f:lambdacase2a}) and
(\ref{f:lambdapmcase2b}). Based on these eigenvalue formulae, linear
stability of solitary waves near pitchfork bifurcations is then
determined. We have shown that the smooth solution branch $u^0(\x;
\mu)$ always switches stability at the bifurcation point. In
addition, the bifurcated solution branches $u^\pm(\x; \mu)$ and the
smooth branch have opposite (same) stability when their power slopes
$P_0'(\mu_0)$ and $P_\pm'(\mu_0)$ have the same (opposite) sign. One
unusual feature of these pitchfork bifurcations in the GNLS
equations is that the smooth and bifurcated solution branches can be
both stable or both unstable, which contrasts such bifurcations in
finite-dimensional dynamical systems where the smooth and bifurcated
branches generally have opposite stability \cite{GH}. For the
special case of positive solitary waves, strong and very explicit
stability results have also been obtained. We have shown that for
positive solitary waves, their linear stability near a pitchfork
bifurcation point can be read off directly from their power diagram.
Lastly, a number of numerical examples of pitchfork bifurcations in
Eq. (\ref{e:U}) have been presented, and the numerical results fully
support the analytical predictions both qualitatively and
quantitatively.

\section*{Acknowledgment}
This work is supported in part by the Air Force Office of Scientific
Research (USAF 9550-09-1-0228) and the National Science Foundation
(DMS-0908167).

\bigskip

\end{document}